\PassOptionsToPackage{usenames, dvipsnames, svgnames, table}{xcolor}
\documentclass[letterpaper,twocolumn,10pt]{article}
\usepackage{usenix-2020-09}

\usepackage{times}
\usepackage{balance}
\usepackage{booktabs} %
\usepackage{xspace}
\usepackage{multirow}
\usepackage{comment}
\usepackage{fancybox, fancyvrb, calc}
\usepackage[center, tight]{subfigure}
\usepackage{enumitem}
\usepackage{amsmath, amssymb}
\usepackage{url}
\usepackage{cite}
\usepackage{breakurl}
\usepackage{graphicx}
\usepackage{pbox}
\usepackage{algorithm, algorithmicx}
\usepackage[noend]{algpseudocode}
\usepackage{amsthm}
\usepackage{balance}
\usepackage{wrapfig}
\usepackage{tikz, pgfplots}
\usetikzlibrary{patterns,matrix,positioning,shadows,backgrounds,arrows,calc,fit,automata,shapes.geometric,arrows,decorations.pathreplacing,decorations.markings,shadings,shapes.symbols,arrows.meta}
\usepackage[capitalise]{cleveref}
\usepackage{tabularx}
\usepackage[T1]{fontenc}
\usepackage[utf8]{inputenc}
\usepackage[normalem]{ulem}
\usepackage{wasysym}
\usepackage{sidecap}
\usepackage{caption}
\usepackage{pifont}
\usepackage[mathcal]{euscript}
\usepackage{setspace}
\usepackage{listings}
\usepackage{footnote}
\usepackage{color,soul}
\makesavenoteenv{tabular}
\makesavenoteenv{table}

\usepackage{hyperref}
\hypersetup{
  colorlinks=true,      %
  linkcolor=blue,       %
  citecolor=magenta,    %
  filecolor=cyan,       %
  urlcolor=blue          %
}
\definecolor{listinggray}{gray}{0.9}
\definecolor{lbcolor}{rgb}{0.9,0.9,0.9}
\usepackage{etoolbox}

\usepackage{bbm}
\usepackage{bm}
\theoremstyle{plain}

\newtheorem{theorem}{Theorem}
\newtheorem{lemma}{Lemma}

\definecolor{boxclr}{gray}{0.9}

\pgfdeclarelayer{bg}    %
\pgfsetlayers{bg,main}  %

\makeatletter
\newcommand{\thickhline}{%
    \noalign {\ifnum 0=`}\fi \hrule height 0.8pt
    \futurelet \reserved@a \@xhline
}
\newcolumntype{"}{@{\vrule width 0.8pt}}
\newcolumntype{[}{@{\vrule width 0.8pt\hskip\tabcolsep}}
\newcolumntype{]}{@{\hskip\tabcolsep\vrule width 0.8pt}}
\newcolumntype{!}{@{\hskip\tabcolsep\vrule width 0.8pt\hskip\tabcolsep}}
\makeatother

\newcommand{\cppsnippet}[1]{%
  \begin{lstlisting}[gobble=4]
    #1
  \end{lstlisting}
}

\tikzset{
    position/.style args={#1:#2 from #3}{
        at=(#3.#1), anchor=#1+180, shift=(#1:#2)
    }
}

\tikzset{
  half fill/.style 2 args={fill=#2, path picture={
    \fill[#1, sharp corners] (path picture bounding box.west) --
                         (path picture bounding box.east) --
                         (path picture bounding box.south east) --
                         (path picture bounding box.south west) -- cycle;}},
}

\tikzset{
  nil fill/.style 2 args={fill=#2, path picture={
    \fill[#1, sharp corners] (path picture bounding box.155) --
                         (path picture bounding box.25) --
                         (path picture bounding box.south east) --
                         (path picture bounding box.south west) -- cycle;}},
}

\tikzset{
  almost fill/.style 2 args={fill=#2, path picture={
    \fill[#1, sharp corners] (path picture bounding box.205) --
                         (path picture bounding box.335) --
                         (path picture bounding box.south east) --
                         (path picture bounding box.south west) -- cycle;}},
}

\definecolor{oldcolor}{HTML}{c66541}

\algdef{SE}[DOWHILE]{Do}{doWhile}{\algorithmicdo}[1]{\algorithmicwhile\ #1}%

\newenvironment{denseitemize}{
\begin{itemize}[topsep=2pt, partopsep=0pt, leftmargin=*]
  \setlength{\itemsep}{2pt}
  \setlength{\parskip}{0pt}
  \setlength{\parsep}{0pt}
}{\end{itemize}}

\newcommand{\cut}[1]{}

\newcommand{\paragraphb}[1]{\vspace{0.075in}\noindent{\bf #1.}}

\cut{
\usepackage{draftwatermark}
\SetWatermarkText{DRAFT -- DO NOT REDISTRIBUTE \ \ \ \ \ \ \ DRAFT -- DO NOT REDISTRIBUTE}
\SetWatermarkFontSize{1.5cm}
\SetWatermarkAngle{307}
\SetWatermarkLightness{0.85}
}

\definecolor{darkgreen}{rgb}{0.1, 0.14, 0.13}

\tikzset{ 
table/.style={
  matrix of nodes,
  row sep=-\pgflinewidth,
  column sep=-\pgflinewidth,
  nodes={rectangle,thick,draw=black,text width={},align=center,font=\small},
  text depth=0.25ex,
  text height=1.25ex,
  nodes in empty cells
},
map/.style={
  matrix of nodes,
  row sep=-\pgflinewidth,
  column sep=-\pgflinewidth,
  nodes={rectangle,draw=black,text width=5em,align=center,font=\small},
  text depth=0.25ex,
  text height=1.25ex,
  nodes in empty cells
},
bigmap/.style={
  matrix of nodes,
  row sep=-\pgflinewidth,
  column sep=-\pgflinewidth,
  nodes={rectangle,draw=black,text width=26em,align=center,font=\small},
  text depth=0.25ex,
  text height=1.25ex,
  nodes in empty cells
},
memcell/.style={
  draw, 
  very thick, 
  text width=0.25em, 
  text height=0.25em
},
}

\tikzstyle{startstop} = [rectangle, rounded corners, minimum width=3em, minimum height=1em,text centered, draw=black, fill=red!30]
\tikzstyle{io} = [trapezium, trapezium left angle=70, trapezium right angle=120, minimum width=2.5em, minimum height=1em, text centered, draw=black, fill=blue!30]
\tikzstyle{process} = [rectangle, minimum width=1.5em, minimum height=1em, align=center, draw=black, fill=gray!30]
\tikzstyle{decision} = [diamond, minimum width=3em, minimum height=1em, align=center, draw=black, fill=SkyBlue!30]
\tikzstyle{arrow} = [thick,->,>=stealth]
\tikzstyle{monolog} = [fill=SkyBlue!30]

\tikzset{
  on each segment/.style={
    decorate,
    decoration={
      show path construction,
      moveto code={},
      lineto code={
        \path [#1]
        (\tikzinputsegmentfirst) -- (\tikzinputsegmentlast);
      },
      curveto code={
        \path [#1] (\tikzinputsegmentfirst)
        .. controls
        (\tikzinputsegmentsupporta) and (\tikzinputsegmentsupportb)
        ..
        (\tikzinputsegmentlast);
      },
      closepath code={
        \path [#1]
        (\tikzinputsegmentfirst) -- (\tikzinputsegmentlast);
      },
    },
  },
  mid arrow/.style={postaction={decorate,decoration={
        markings,
        mark=at position .5 with {\arrow[#1]{stealth}}
      }}},
}

\algnewcommand{\IIf}[1]{\State\algorithmicif\ #1\ \algorithmicthen}%
\algnewcommand{\EndIIf}{\unskip\ }%
\algdef{SE}[DOWHILE]{Do}{doWhile}{\algorithmicdo}[1]{\algorithmicwhile\ #1}%
\algnewcommand\algorithmicforeach{\textbf{for each}}%
\algdef{S}[FOR]{ForEach}[1]{\algorithmicforeach\ #1\ \algorithmicdo}%

\newcounter{mynote}[section]

\def\eg{{\em e.g.}\xspace}

\def\name{Karma\xspace}
\def\mp{karmaPool\xspace}
\def\credit{credit\xspace}

\newcommand{\One}[1]{\mathbbm{1}\left[#1\right]}
\usepackage{graphicx}
\usepackage{epstopdf}
\usepackage[export]{adjustbox}
\usepackage[subtle]{savetrees}
\begin{document}

\date{}

\title{\name: Resource Allocation for Dynamic Demands}

\author{
\hspace{0.25in}{\rm Midhul Vuppalapati} \hspace{0.15in}
\\\hspace{0.25in}{Cornell University} \hspace{0.15in}
\and
{\rm Giannis Fikioris} \hspace{0.15in}
\\{Cornell University} \hspace{0.15in}
\and
{\rm Rachit Agarwal} \hspace{0.2in}
\\{Cornell University} \hspace{0.2in}
\and
{\rm Asaf Cidon} \hspace{0.05in}
\\{Columbia University} \hspace{0.05in}
\and
{\rm Anurag Khandelwal} \hspace{0.05in}
\\{Yale University} \hspace{0.05in}
\and
{\rm \'Eva Tardos} \hspace{-0.0in}
\\{Cornell University} \hspace{0.0in}
}

\maketitle

\begin{sloppypar}
\begin{abstract}
\vspace{0.05in}
\noindent
We consider the problem of fair resource allocation in a system where user demands are dynamic, that is, where user demands vary over time. Our key observation is that the classical max-min fairness algorithm for resource allocation provides many desirable properties (\eg, Pareto efficiency, strategy-proofness, and fairness), but only under the strong assumption of user demands being static over time. For the realistic case of dynamic user demands, the max-min fairness algorithm loses one or more of these properties.

We present \name, a new resource allocation mechanism for dynamic user demands. The key technical contribution in \name is a credit-based resource allocation algorithm: in each quantum, users donate their unused resources and are assigned credits when other users borrow these resources; \name carefully orchestrates the exchange of credits across users (based on their instantaneous demands, donated resources and borrowed resources), and performs prioritized resource allocation based on users' credits. We theoretically establish \name guarantees related to Pareto efficiency, strategy-proofness, and fairness for dynamic user demands. Empirical evaluations over production workloads show that these properties translate well into practice: \name is able to reduce disparity in performance across users to a bare minimum while maintaining Pareto-optimal system-wide performance.

\end{abstract}
  \vspace{-0.1in}
\section{Introduction}
\label{sec:intro}
  \vspace{-0.05in}
Resource allocation is a fundamental problem in computer systems, spanning private and public clouds, computer networks, hypervisors, etc. There is a large and active body of research on designing resource allocation mechanisms that achieve Pareto efficiency (high resource utilization) and strategy-proofness (selfish users should not be able to benefit by lying about their demands) while ensuring that resources are allocated fairly among users, \eg,~\cite{carbyne, pisces, drf, stoica1996proportional, fairride, quincy, faircloud}.

For a system containing a single resource, the two most popular allocation mechanisms are strict partitioning~\cite{snowset,atikoglu2012workload} and max-min fairness~\cite{carbyne,drf,pisces,faircloud,fairride,srikanth-pop, srikanth-swan, b4, gavel}. The former allocates the resource equally across all users (``fair share''), independent of their demands; this guarantees strategy-proofness and fairness, but not Pareto efficiency since resources can be underutilized when one or more users have demands lower than the fair share. Max-min fairness alleviates limitations of strict partitioning by taking user demands into account: it maximizes the minimum allocation across users while ensuring that each user's allocation is no more than their demand. A classical result shows that resource allocation based on max-min fairness guarantees each of the three desirable properties---Pareto efficiency, strategy-proofness, and fairness. These powerful properties have, over decades, motivated efforts in both systems and theory communities on generalizations of max-min fairness for allocating multiple resources~\cite{drf, tetris, carbyne}, for incorporating application performance goals and deadlines~\cite{quincy, tetris, themis, dpf}, and for new models of resource allocation~\cite{pisces, fairride, graphene, hug, ltf1, feigenbaum2004distributed}, to name a few.

This paper explores a complementary problem---resource allocation of a single elastic resource in a system where user demands are dynamic, that is, vary over time. Dynamic user demands are the norm in most real-world deployments~\cite{googletrace, alibabatrace1, alibabatrace2, snowset, twitter-caching, jiffy, berg2020cachelib, shahrad2020serverless, borg}; for instance, analysis of production workloads in \S\ref{sec:overview} reveals that user demands vary by as much as $17\times$ within minutes, with majority of users having demands with standard deviation $0.5-43\times$ of the average over time. We show in \S\ref{sec:overview} that, for systems with such dynamic user demands, resource allocation based on the max-min fairness algorithm fails to guarantee one or more of its properties: (1) if the allocation is done based on demands at $t=0$, Pareto efficiency and strategy-proofness are no longer guaranteed; and, (2) if the allocation is done periodically, {\em long-term} fairness is no longer guaranteed---for $n$ users with the same average demand, the max-min fairness algorithm may allocate some user as much as $\Omega(n)$ more resources than other users over time. %

We present \name, a new resource allocation mechanism for dynamic user demands. The key technical contribution of \name is a credit-based resource allocation algorithm: in each quantum, users receive credits when they donate a part of their fair share of resources (\eg, if their demand is less than their fair share); users can use these credits to borrow resources in any future quantum when their demand is higher than their fair share. When the supply of resources from donors is equal to the demand from borrowers, it is easy to exchange resources and credits among users. The key algorithmic challenge that \name resolves is when supply is not equal to demand---in such scenarios, \name carefully orchestrates resources and credits between donors and borrowers: donors are prioritized so as to keep credits across users as balanced as possible, and borrowers are prioritized so as to keep the resource allocation as fair as possible. 

We theoretically establish \name guarantees for dynamic user demands. \name guarantees Pareto efficiency at all times: in each quantum, it allocates resources such that it is not possible to increase the allocation of a user without decreasing the allocation of at least another user. For strategy-proofness, \name guarantees that a selfish user cannot increase their aggregate resource allocation by {\em over}-reporting their demands in any quantum. In addition, we show a new surprising phenomenon (that may primarily be of theoretical interest): if a user had perfect knowledge about the future demands of all other users, the user can increase its own aggregate allocation by a small constant factor by {\em under}-reporting its demand in some quanta; however, for $n$ users, imprecision in this future knowledge could lead to the user losing $\Omega(n)$ factor of their aggregate resource allocation by under-reporting their demand in any quantum. Put together, these results enable \name to provide powerful guarantees related to strategy-proofness. Finally, for fairness, we prove that given a set of (past) allocations, \name guarantees an optimally-fair resource allocation. We also establish that \name guarantees similar properties even when multiple selfish users can collude, and even when different users have different fair shares.

We have realized \name on top of Jiffy~\cite{jiffy}, an open-sourced multi-tenant elastic memory system; an end-to-end implementation of \name is available at \url{https://github.com/resource-disaggregation/karma}. 
Evaluation of \name over production workloads demonstrates that \name's theoretical guarantees translate well into practice: it matches the max-min fairness algorithm in terms of resource utilization, while significantly improving the long-term fairness of resources allocated across users. \name's fairer resource allocation directly translates to application-level performance; for instance, over evaluated workloads, \name keeps the {\em average} performance (across users) the same as the max-min fairness algorithm, while reducing performance {\em disparity} across users by as much as ${\sim}2.4\times$. \name also incentivizes users to share resources: our evaluation shows that (1) \name-conformant users achieve much more desirable allocation and performance compared to users who prefer a dedicated fair share of resources; and, (2) if users were to turn \name-conformant, they can improve their performance by better matching their allocations with their demands over time.

  \vspace{-0.1in}
\section{Motivation}
\label{sec:overview}
  \vspace{-0.05in}
We begin by outlining our motivating use cases, followed by an in-depth discussion on the limitations of the classic max-min fairness algorithm for dynamic user demands.  

\begin{figure*}[t]
  \centering
  \includegraphics[width=0.25\textwidth]{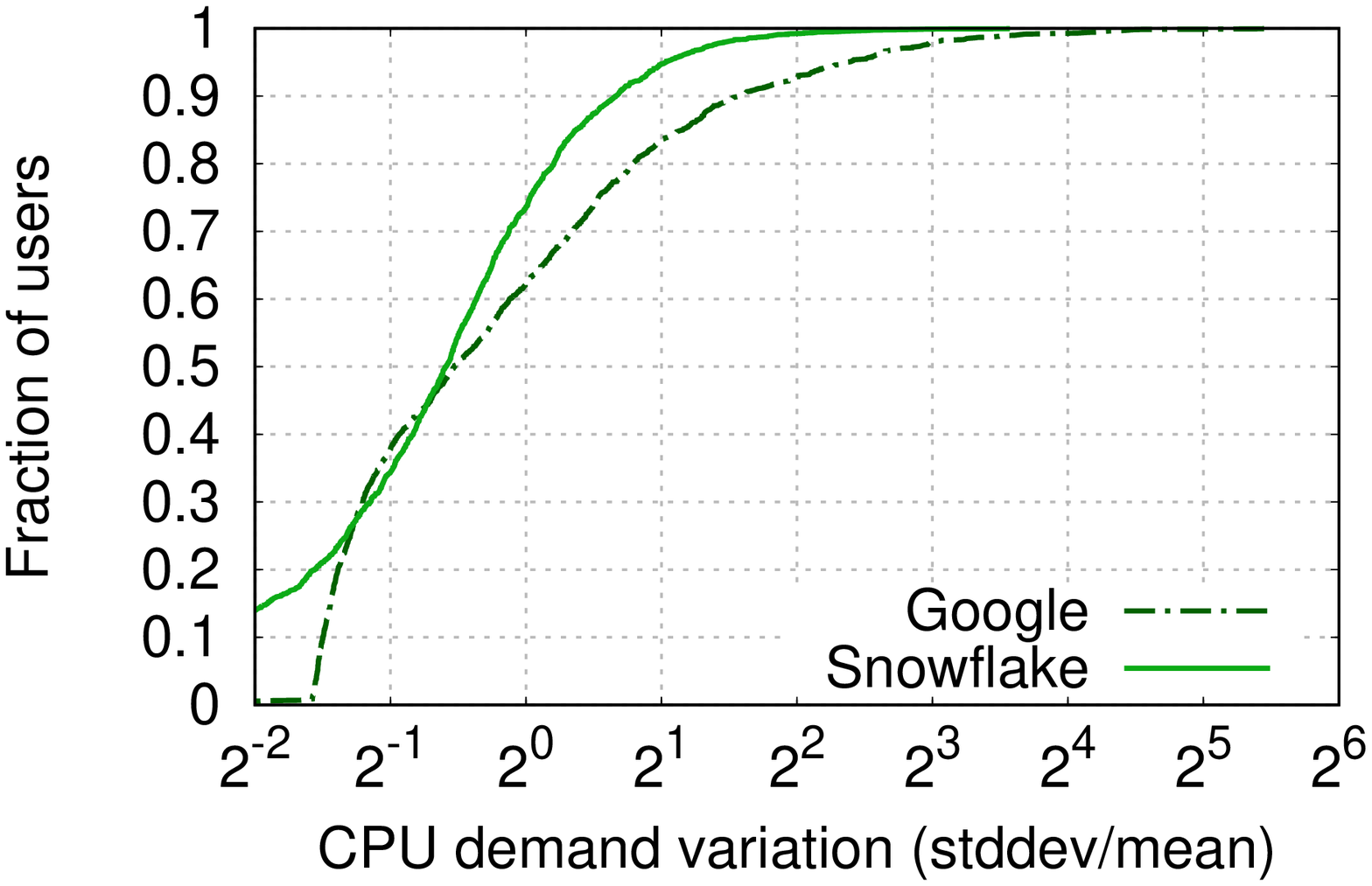}
  \includegraphics[width=0.25\textwidth]{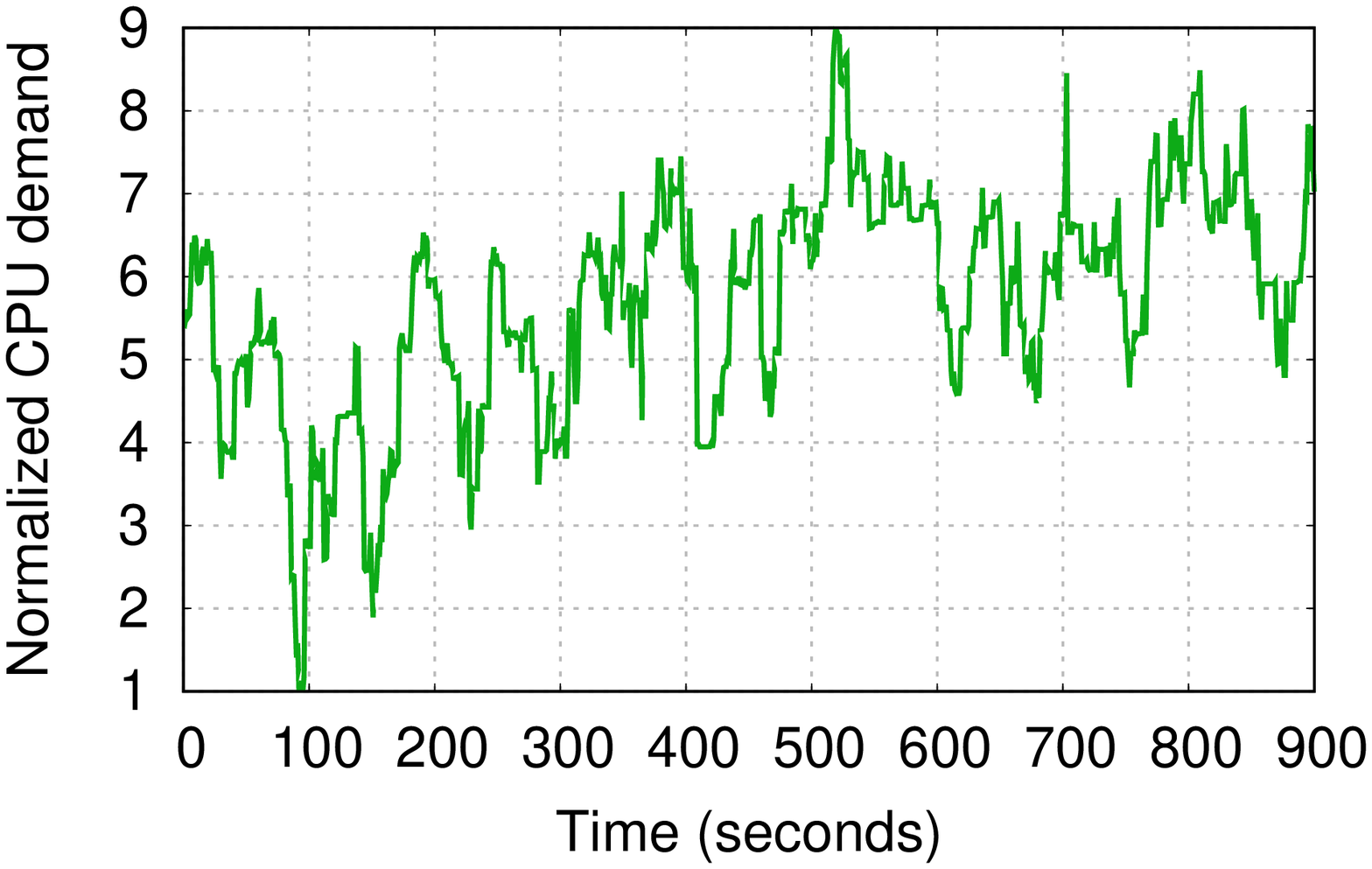}
  \includegraphics[width=0.25\textwidth]{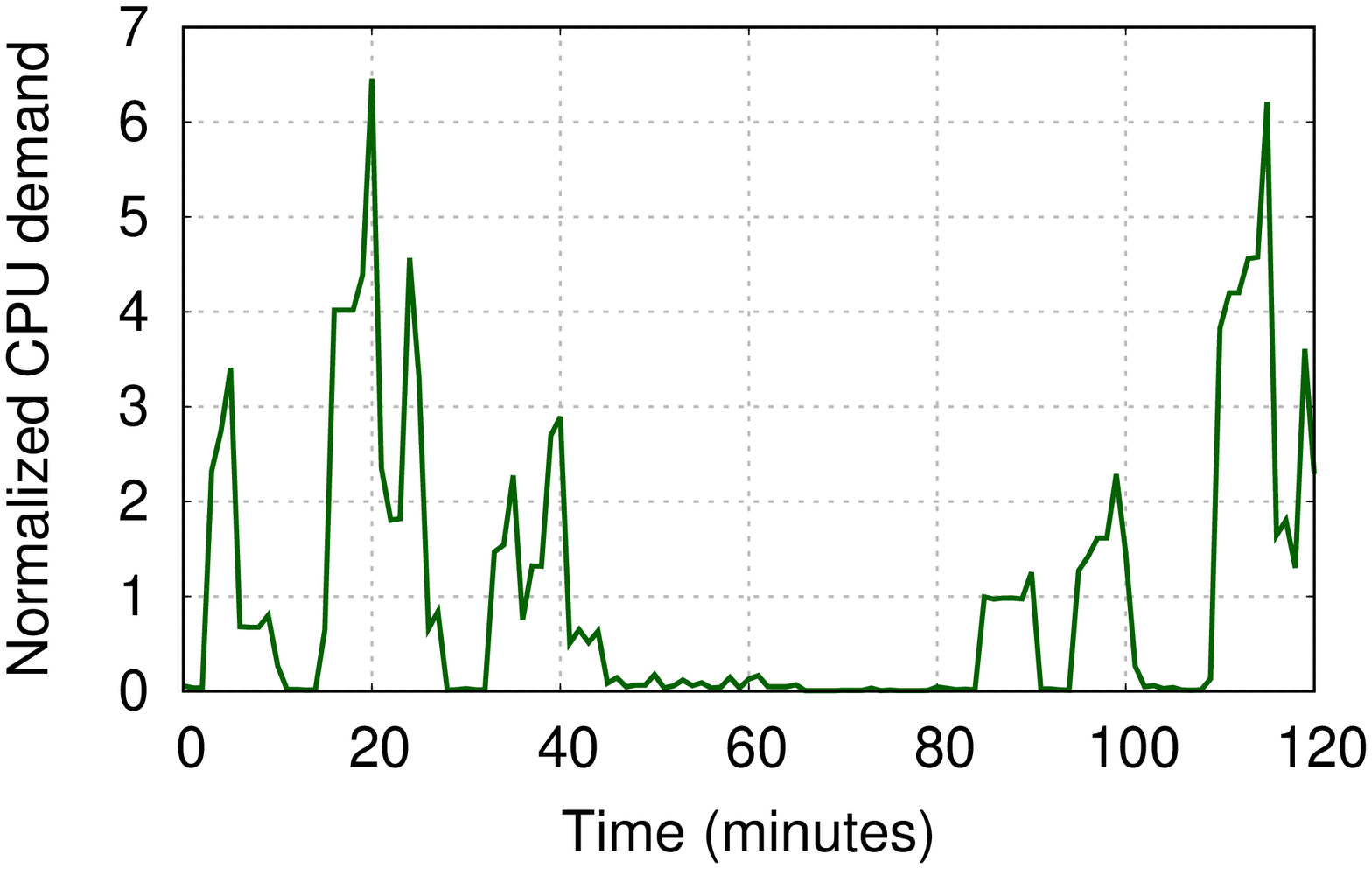}
  
  \includegraphics[width=0.25\textwidth]{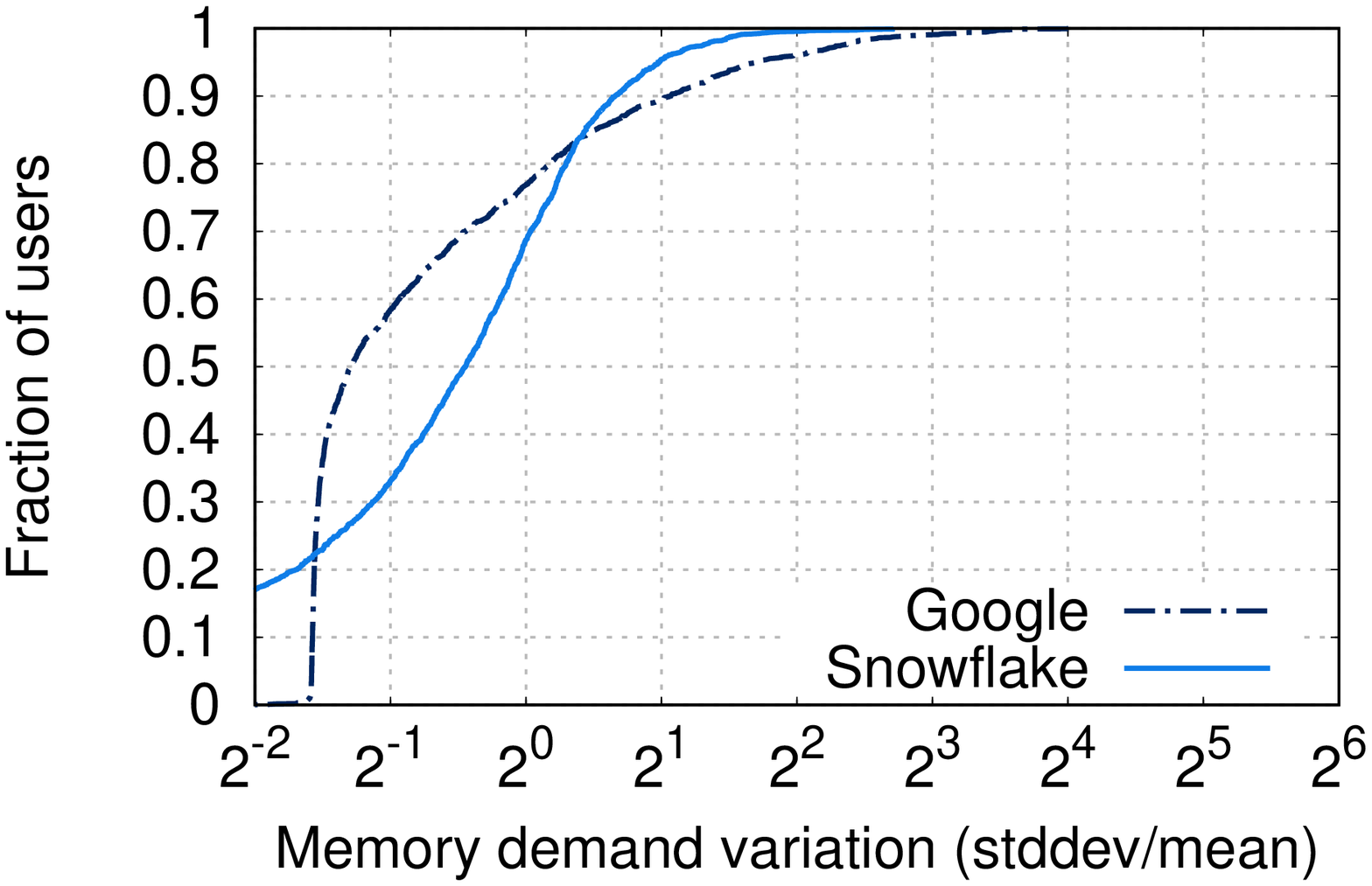}
  \includegraphics[width=0.25\textwidth]{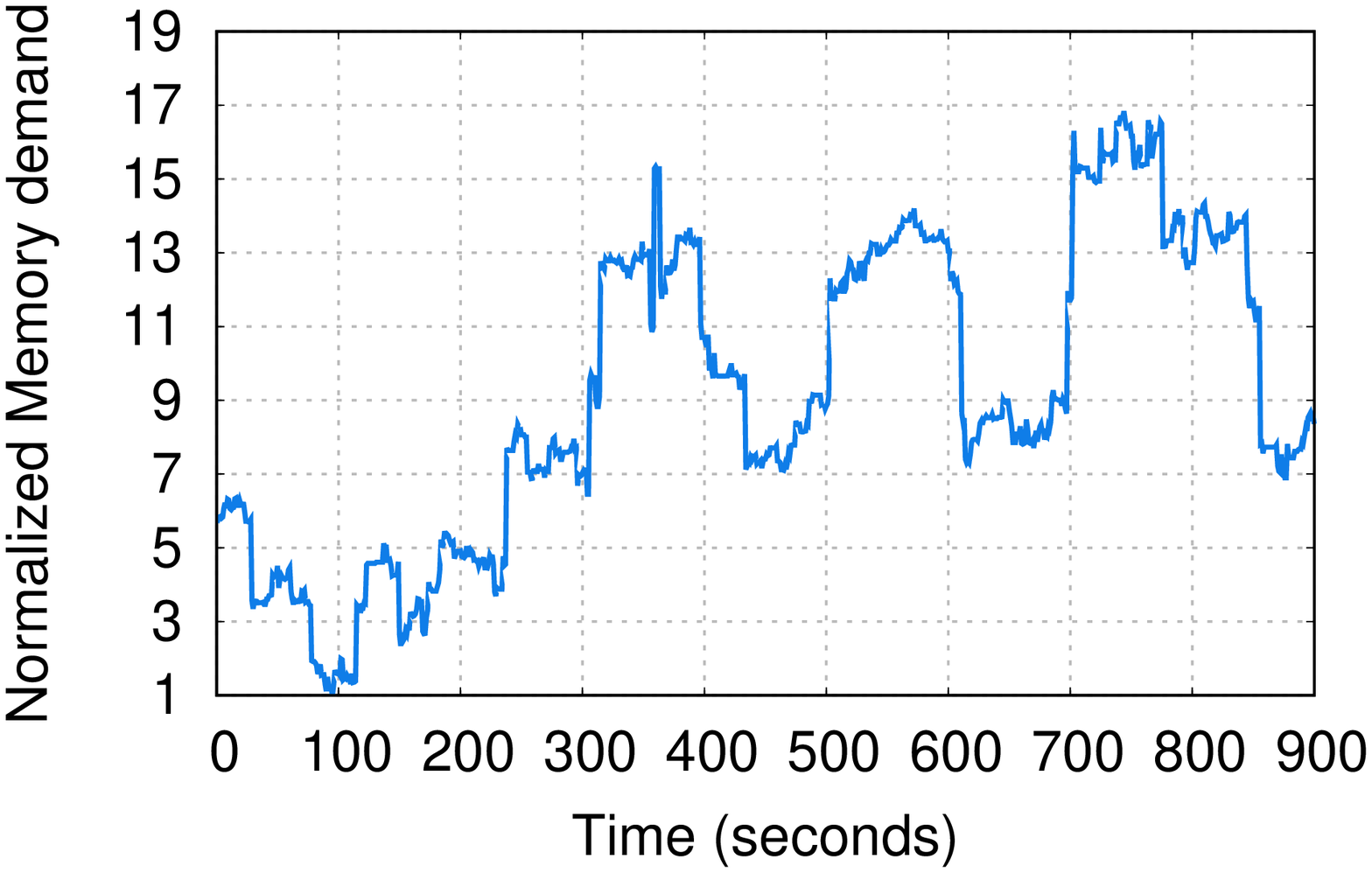}
  \includegraphics[width=0.25\textwidth]{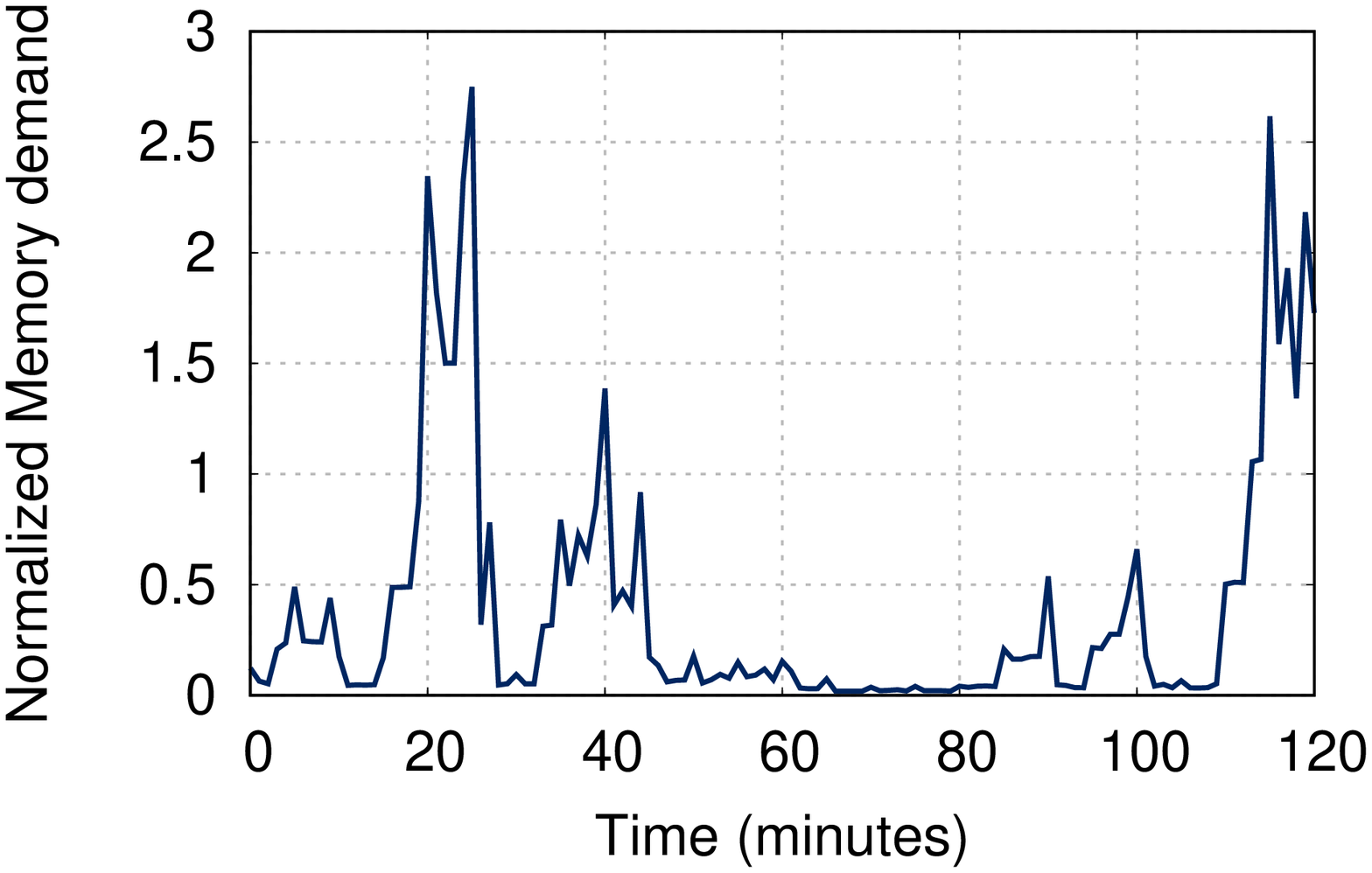}
  \vspace{-0.1in}
  \caption{
    \textbf{Analysis of Google and Snowflake workloads suggests that a large fraction of users have dynamic demands (left) that can change dramatically over short timescales (center, right)} (Left) CDFs, across users, of the ratio of standard deviation and mean of each user's demand. (Center) For a randomly sampled user in the Snowflake trace, the variation in the user's CPU and memory demands (normalized by minimum demand) over a $15$ minute period. (Right) For a randomly sampled user in the Google trace, the variation in the user's CPU and memory demands (normalized by minimum demand) over a $2$ hour period.
    \label{fig:motivation}
    }
  \vspace{-0.1in}
  \end{figure*}

\paragraphb{Motivating use cases}
Fair resource allocation is an important problem in private clouds where resources are shared by multiple users or teams within the same organization~\cite{graphene, carbyne, drf, tetris, pisces, hug, dpf, alibabatrace1, googletrace, alibabatrace2, fairride, delaysc, quincy, snowset, berg2020cachelib, borg, twitter-caching, b4, srikanth-swan}; our primary use cases are from such private clouds. \name may also be useful for emerging use cases from multi-tenant public clouds where spare resources may be allocated to tenants while providing performance isolation~\cite{blk-switch, netchannel, seawall,faircloud,pulsar,pisces, jiffy, shahrad2020serverless}. We discuss motivating scenarios in both contexts below.

One scenario is shared analytics clusters. For instance, companies like Microsoft, Google, and Alibaba employ schedulers~\cite{borg, yarn, mesos, delaysc, quincy, carbyne} that allocate resources across multiple internal teams that run long-running jobs (\eg, for data analytics~\cite{mapreduce, spark}) on a shared set of resources. Consider memory as a shared resource; in many of these frameworks, main memory is used to cache frequently accessed data from slower persistent storage and to store intermediate data generated during job execution. Indeed, increasing the allocated memory improves job performance; however, since memory is limited and is shared across multiple teams, ensuring resource allocation fairness is also a key requirement. Moreover, since these jobs are usually long-running, their performance depends on long-term memory allocations, rather than instantaneous allocations~\cite{carbyne, alibabatrace1, alibabatrace2}.

Another use case is shared caches: many companies (\eg Facebook~\cite{scalingmemcached,berg2020cachelib,atikoglu2012workload} and Twitter~\cite{twitter-caching}) operate clusters of in-memory key-value caches, such as memcached or Redis, serving a wide array of internal applications. In this use case, the memory demand of each application may be computed as the amount of memory that would be required to fit hot objects within the cache~\cite{scalingmemcached, memshare, twitter-caching, cliffhanger}. In such settings, efficient and fair sharing of caches is of utmost importance~\cite{memshare, snowset, scalingmemcached,atikoglu2012workload}: to maintain service level agreements, it is important to have consistently good performance over long periods of time, rather than excellent performance at some times and very poor performance at other times (see~\cite{memshare, snowset, scalingmemcached,atikoglu2012workload} for more discussion on the importance of long-term performance). 

Third, fair resource allocation while ensuring high utilization is also a goal in inter-datacenter bandwidth allocation~\cite{srikanth-pop, srikanth-swan, b4}. Existing traffic engineering solutions used in production environments perform periodic max-min fair resource allocation to account for dynamic user demands~\cite{srikanth-pop, srikanth-swan, b4}. Our work demonstrates that periodically performing max-min fair resource allocation over such dynamic demands leads to unfair resource allocation across users.

Finally, an interesting use case in the public cloud context is that of burstable VMs~\cite{aws-burstable,azure-burstable} that use virtual currency to enable resource allocation over dynamic user demands. These VMs share resources with VMs from other users and are charged on an instance-specific baseline. When resource utilization is below the baseline, users accumulate virtual currency that they can later use to gain resources beyond the baseline during periods of high demand. Given that Burstable VMs are primarily useful for dynamic user demands, they will likely need resource allocation mechanisms that guarantee high utilization, strategy-proofness, and fair resource allocation.

\paragraphb{Dynamic user demands}
Increasingly many applications running data analytics or key-value caches operate on data collected from social media, application and network logs, mobile systems, etc. A unique characteristic of these data is that they are less controllable by the organization because they are generated by entities outside of the organization. As a result, applications can observe highly time-varying dynamic resource demands~\cite{googletrace, alibabatrace1, alibabatrace2, snowset, twitter-caching, jiffy, berg2020cachelib, shahrad2020serverless, borg}.

To build a deeper understanding of variation in user demands over time, we analyze two publicly-available production workloads: (1) Google~\cite{googletrace} resource usage information across $8$ clusters ($1000-2000$ users per cluster) over a $30$ day period; and, (2) Snowflake~\cite{snowset}, a cloud-based database query engine that provides resource usage statistics for over $2000$ users over a $14$ day period. To characterize user demand variability over time, we compute---for each user---the ratio of the standard deviation and mean of their demands over the entire period. Figure~\ref{fig:motivation}~(left) shows that $40-70$\% of all users in both Google and Snowflake workloads have a standard deviation in CPU and memory demands at least $0.5\times$ their mean, indicating high variability in demands for most users. Furthermore, the standard deviation in demands of as many as $20\%$ of the users can be as high as their mean demand, with some users having extremely high variance in demands (standard deviations up to $12-43\times$ the mean). Similar observations have been made for time-varying user demands in inter-datacenter networks; for instance, production studies~\cite{ncflow} show that, on average, user demands vary by $35\%$ within $5$-minute intervals, with some demands varying by as much as $45\%$ within a short period of time.

Figure~\ref{fig:motivation}~(center) shows the CPU and memory demands for a randomly-sampled user from the Snowflake trace over a $15$ minute window (we show only one user and only $15$ minute window for clarity; analyzing a sample of $100$ users, we find $87\%$ of the users to have similar demand patterns). The figure shows that user demands can change dramatically over tens of seconds, by as much as $6\times$ and $2\times$ for compute and memory, respectively. Similarly, we see significant variation in demands even for a random user from the Google trace (shown in Figure~\ref{fig:motivation}~(right)).

\paragraphb{Max-min fairness guarantees fail for dynamic user demands}
The classical max-min fairness algorithm for resource allocation provides many desirable properties, \eg, Pareto efficiency, strategy-proofness, and fairness. However, buried under the proofs is the assumption that user demands are static over time, an assumption that does not hold in practice (as demonstrated in Figure~\ref{fig:motivation}). For the realistic case of dynamic user demands, max-min fairness can be applied in two ways, each of which leads to violating one or more of its properties. We will demonstrate this using the example in Figure~\ref{fig:max-min-example}; here, time is divided into five quanta and three users have demands varying across quanta. %

First, one can na\"ively perform max-min fair allocation just once based on user demands at quantum $t=0$. This results in max-min fairness losing both Pareto efficiency and strategy-proofness. In the example of Figure~\ref{fig:max-min-example}, since allocations will only be done based on the demands specified by the users at $t=0$, if users were to specify their true demands, user {\textrm C} will obtain an allocation of $1$ unit leading to a total useful allocation of $3$ units over the entire duration (as shown in Figure~\ref{fig:max-min-example}~(middle, top)); if user C were to lie and over-report their demand at $t=0$ as $2$ units, then they can achieve a more desirable total useful allocation of $5$ units (Figure~\ref{fig:max-min-example}~(middle, bottom)). This breaks strategy-proofness. In addition, max-min fairness is also not Pareto efficient: for many quanta, resources allocated to users will be underutilized as is evident in Figure~\ref{fig:max-min-example}~(middle).  

\tikzset{solidbox/.style={draw=black, thick}}
\definecolor{nicered}{HTML}{5184D9}
\definecolor{niceblue}{HTML}{D95184}
\definecolor{nicegreen}{HTML}{84D951}
\tikzset{C/.style={fill=nicered!85}, A/.style={preaction={fill,niceblue!85}, pattern=north east lines}, B/.style={fill=nicegreen!85}}
\tikzset{CU/.style={fill=white}, AU/.style={fill=white}, BU/.style={fill=white}}
\tikzset{D/.style={fill=yellow!85}}
\tikzset{legenditem/.style={draw=black, thick, inner sep=0, minimum width=4mm, minimum height=3mm}}
\tikzset{pics/stackedbar/.style n args={7}{
    code = {
        \filldraw[#1, #2] (0, 0) rectangle (1, #5);
        \filldraw[shift={(0, #5)}] [#1, #3] (0, 0) rectangle (1, #6);
        \filldraw[shift={(0, #5 + #6)}] [#1, #4] (0, 0) rectangle (1, #7);
    }
}}
\tikzset{pics/stackedbar4/.style n args={9}{
  code = {
      \filldraw[#1, #2] (0, 0) rectangle (1, #6);
      \filldraw[shift={(0, #6)}] [#1, #3] (0, 0) rectangle (1, #7);
      \filldraw[shift={(0, #6 + #7)}] [#1, #4] (0, 0) rectangle (1, #8);
      \filldraw[shift={(0, #6 + #7 + #8)}] [#1, #5] (0, 0) rectangle (1, #9);
  }
}}
\tikzset{pics/stackedbar6/.style n args={6}{
  code = {
      \filldraw[solidbox, A] (0, 0) rectangle (1, #1);
      \filldraw[shift={(0, #1)}] [solidbox, AU] (0, 0) rectangle (1, #2);
      \filldraw[shift={(0, #1 + #2)}] [solidbox, B] (0, 0) rectangle (1, #3);
      \filldraw[shift={(0, #1 + #2 + #3)}] [solidbox, BU] (0, 0) rectangle (1, #4);
      \filldraw[shift={(0, #1 + #2 + #3 + #4)}] [solidbox, C] (0, 0) rectangle (1, #5);
      \filldraw[shift={(0, #1 + #2 + #3 + #4 + #5)}] [solidbox, CU] (0, 0) rectangle (1, #6);
  }
}}
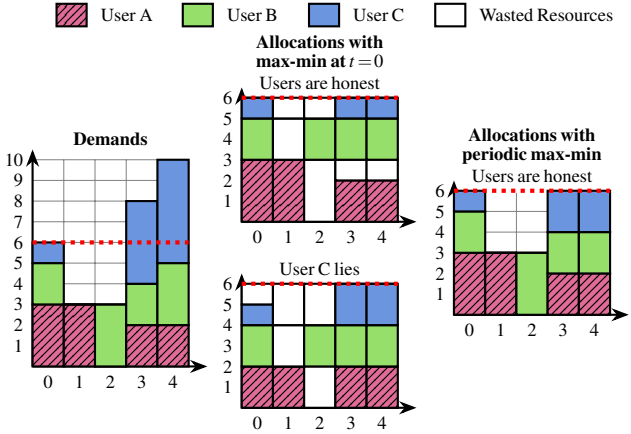
\begin{figure}
  \centering
    \begin{tikzpicture}[xscale=0.415, yscale=0.275, font=\scriptsize, >=Stealth]

      \begin{scope}[shift={(-3.5, 0)}]
      \draw[step=1,gray,very thin] (0, 0) grid (5, 10);
      \draw[->, thick] (0, 0) -- (5.6, 0);
      \draw[->, thick] (0, 0) -- (0, 10.5);
      \foreach \x/\xtext in {0.5/0, 1.5/1, 2.5/2, 3.5/3, 4.5/4}
          \node[below] at (\x, 0) {$\xtext$};
      \foreach \y in {1, 2,..., 10}
          \node[left] at (0, \y) {$\y$};
          
      \draw (0,0) pic[transform shape] {stackedbar={solidbox}{A}{B}{C}{3}{2}{1}};
      \draw[shift={(1,0)}] (0,0) pic[transform shape] {stackedbar={solidbox}{A}{B}{C}{3}{0}{0}};
      \draw[shift={(2,0)}] (0,0) pic[transform shape] {stackedbar={solidbox}{A}{B}{C}{0}{3}{0}};
      \draw[shift={(3,0)}] (0,0) pic[transform shape] {stackedbar={solidbox}{A}{B}{C}{2}{2}{4}};
      \draw[shift={(4,0)}] (0,0) pic[transform shape] {stackedbar={solidbox}{A}{B}{C}{2}{3}{5}};
      
      \draw[dotted, red, ultra thick] (0, 6) -- (5, 6); 
      
      \node at (2.5, 11) {\textbf{Demands}};
      \end{scope}

      \begin{scope}[shift={(3.2, -2)}]
      \draw[step=1,gray,very thin] (0, 0) grid (5, 5.9);
      \draw[->, thick] (0, 0) -- (5.6, 0);
      \draw[->, thick] (0, 0) -- (0, 6.5);
      \foreach \x/\xtext in {0.5/0, 1.5/1, 2.5/2, 3.5/3, 4.5/4}
          \node[below] at (\x, 0) {$\xtext$};
      \foreach \y in {1, 2,..., 6}
          \node[left] at (0, \y) {$\y$};

      \draw (0,0) pic[transform shape] {stackedbar6={2}{0}{2}{0}{1}{1}};
      \draw[shift={(1,0)}] (0,0) pic[transform shape] {stackedbar6={2}{0}{0}{2}{0}{2}};
      \draw[shift={(2,0)}] (0,0) pic[transform shape] {stackedbar6={0}{2}{2}{0}{0}{2}};
      \draw[shift={(3,0)}] (0,0) pic[transform shape] {stackedbar6={2}{0}{2}{0}{2}{0}};
      \draw[shift={(4,0)}] (0,0) pic[transform shape] {stackedbar6={2}{0}{2}{0}{2}{0}};
      
      \draw[dotted, red, ultra thick] (0, 6) -- (5, 6); 
      
      \node at (2.5, 17.7) {\textbf{Allocations with}};
      \node at (2.5, 16.8) {\textbf{max-min at $t=0$}};
      \node at (2.5, 6.7) {User C lies};
      \end{scope}

      \begin{scope}[shift={(3.2, 7)}]
      \draw[step=1,gray,very thin] (0, 0) grid (5, 5.9);
      \draw[->, thick] (0, 0) -- (5.6, 0);
      \draw[->, thick] (0, 0) -- (0, 6.5);
      \foreach \x/\xtext in {0.5/0, 1.5/1, 2.5/2, 3.5/3, 4.5/4}
          \node[below] at (\x, 0) {$\xtext$};
      \foreach \y in {1, 2,..., 6}
          \node[left] at (0, \y) {$\y$};

      \draw (0,0) pic[transform shape] {stackedbar6={3}{0}{2}{0}{1}{0}};
      \draw[shift={(1,0)}] (0,0) pic[transform shape] {stackedbar6={3}{0}{0}{2}{0}{1}};
      \draw[shift={(2,0)}] (0,0) pic[transform shape] {stackedbar6={0}{3}{2}{0}{0}{1}};
      \draw[shift={(3,0)}] (0,0) pic[transform shape] {stackedbar6={2}{1}{2}{0}{1}{0}};
      \draw[shift={(4,0)}] (0,0) pic[transform shape] {stackedbar6={2}{1}{2}{0}{1}{0}};
      
      \draw[dotted, red, ultra thick] (0, 6) -- (5, 6); 
      
      \node at (2.5, 6.7) {Users are honest};
      \end{scope}

        \begin{scope}[shift={(10, 2.5)}]
        \draw[step=1,gray,very thin] (0, 0) grid (5, 5.9);
        \draw[->, thick] (0, 0) -- (5.6, 0);
        \draw[->, thick] (0, 0) -- (0, 6.5);
        \foreach \x/\xtext in {0.5/0, 1.5/1, 2.5/2, 3.5/3, 4.5/4}
            \node[below] at (\x, 0) {$\xtext$};
        \foreach \y in {1, 2,..., 6}
            \node[left] at (0, \y) {$\y$};

        \draw (0,0) pic[transform shape] {stackedbar={solidbox}{A}{B}{C}{3}{2}{1}};
      \draw[shift={(1,0)}] (0,0) pic[transform shape] {stackedbar={solidbox}{A}{B}{C}{3}{0}{0}};
      \draw[shift={(2,0)}] (0,0) pic[transform shape] {stackedbar={solidbox}{A}{B}{C}{0}{3}{0}};
      \draw[shift={(3,0)}] (0,0) pic[transform shape] {stackedbar={solidbox}{A}{B}{C}{2}{2}{2}};
      \draw[shift={(4,0)}] (0,0) pic[transform shape] {stackedbar={solidbox}{A}{B}{C}{2}{2}{2}};
        
        \draw[dotted, red, ultra thick] (0, 6) -- (5, 6); 
        
        \node at (2.5, 8.7) {\textbf{Allocations with}};
        \node at (2.5, 7.7) {\textbf{periodic max-min}};
        \node at (2.5, 6.7) {Users are honest};
        \end{scope}

      \begin{scope}[shift={(-2.25, 17)}, node distance=1.25]
      \node (legendA) [legenditem, A, label={[shift={(0.05,0)}]right:{User A}}] at (0, 0) {};
      \node (legendB) [legenditem, B, label={[shift={(0.05,0)}]right:{User B}}, right=of legendA] {};
      \node (legendC) [legenditem, C, label={[shift={(0.05,0)}]right:{User C}}, right=of legendB] {};
      \node (legendC) [legenditem, AU, label={[shift={(0.05,0)}]right:{Wasted Resources}}, right=of legendC] {};
      \end{scope}
      \end{tikzpicture}
  \vspace{-0.1in}
  \caption{\textbf{Classical max-min fairness guarantees break for dynamic user demands.} Here, $6$ units of a resource are shared by $3$ users (fair share of $2$). Discussion in \S\ref{sec:overview}.
  }
  \label{fig:max-min-example}
  \vspace{-0.1in}
\end{figure}

A better way to apply max-min fairness for dynamic user demands is to periodically reallocate resources based on users' instantaneous demands (\eg, every quantum of time periods, as in several operating systems and hypervisors~\cite{cfs, ESX}). This trivially guarantees Pareto efficiency and strategy-proofness but results in extremely unfair allocation across users. Figure~\ref{fig:max-min-example}~(right, top) shows an example where max-min fairness can result in $2\times$ disparity between resources allocated to users over the $5$ quanta---user A receives a total allocation of $10$ slices, while user C receives a total allocation of only $5$ slices, despite them having the same average demand; this example can be easily extended to demonstrate that max-min fairness can, for $n$ users, result in resource allocations where some user gets a factor of $\Omega(n)$ larger amount of resources than other users (proof in \S\ref{max-min-bad}). Such disparity in resource allocations also leads to disparity in application-level performance across users since, as discussed above in use cases, many applications require consistently good performance over long periods of time, rather than excellent performance at some times and very poor performance at other times~\cite{carbyne, ltf1, ltf2, ltf3}. We will demonstrate, in the evaluation section, that users experience significant disparity in application-level performance due to such disparate resource allocations. 

For the rest of the paper, we focus on long-term fairness; informally, an allocation is considered fair if all users have the same aggregate resource allocation over time. Our goal is to design a resource allocation mechanism that, for dynamic user demands, guarantees Pareto efficiency, strategy-proofness, and fairness.

  \vspace{-0.1in}
\section{Karma}
\label{ssec:altruism}
  \vspace{-0.05in}
\name is a resource allocation mechanism for dynamic user demands. \name uses \textit{credits} (\S\ref{ssec:mempool}, \S\ref{ssec:incentive})---users receive credits when they donate a part of their fair share of resources (\eg, when their demand is less than their fair share), and can use these credits to borrow resources beyond their fair share during periods of high demand. \name carefully orchestrates the exchange of resources and credits between donors and borrowers: donors are prioritized in a manner that ensures credit distribution across users remains as balanced as possible, and borrowers are prioritized in a manner that keeps the resource allocation as fair as possible. We will prove theoretically in \S\ref{ssec:formal} that, while simple in hindsight, this allocation mechanism simultaneously achieves Pareto efficiency, strategy-proofness, and fairness for dynamic user demands.

  \vspace{-0.1in}
\subsection{Preliminaries}
\label{ssec:mempool}
  \vspace{-0.05in}
We consider the following setup for the problem: we have $n$ users sharing a single resource (CPU, memory, GPUs, etc.); each user has a fair share of $f$ resource units (each unit is referred to as a {\em slice}), and thus the pool has $n \times f$ slices of the resource (as we discuss in \S\ref{ssec:algo-discussion}, all our results hold for users having different fair shares). Time is divided into quanta, users demand a certain number of resource slices every quantum, and \name performs resource (re)allocation at the beginning of each quantum. While user demands during each quantum can be arbitrary, unsatisfied demands in one quantum do not carry over to the next. Similar to prior work~\cite{drf,pisces,faircloud,fairride}, we assume that users are not adversarial (that is, do not lie about their demands simply to hurt others' allocations), but are otherwise selfish and strategic (willing to misreport their demands to maximize their allocations).

\vspace{-0.1in}
\subsection{\name design}
\label{ssec:incentive}
  \vspace{-0.05in}
Let $0 \leq \alpha \leq 1$ be a parameter. \name guarantees that each user is allocated an $\alpha$ fraction of its fair share $(= \alpha \cdot f)$ in each quantum; we refer to this as the guaranteed share. 
\name maintains a pool of resource slices---\mp---that, at any point in time, contains two types of slices:
\begin{denseitemize}
\item \textbf{Shared slices} are the slices in the resource pool that are not guaranteed to any user. It is easy to see that the number of shared slices in the system is $n \cdot f - n \cdot \alpha \cdot f = n \cdot (1-\alpha) \cdot f$.
\item \textbf{Donated slices}, that are donated by users whose demands are smaller than their guaranteed share.
\end{denseitemize}

\noindent
We use these two sets of slices in the following manner. In any given quantum, if a user has demand less than its guaranteed share, then the user is said to be ``donating'' as many slices as the difference between the user's guaranteed share and demand in that quantum. A user that has demand larger than its guaranteed share is said to be ``borrowing'' slices beyond its guaranteed share, which the system can potentially supply using either shared slices or donated slices.

\vspace{-0.1in}
\subsubsection{\name credits}
\label{sssec:currency}
\name allocates resources not just based on users' instantaneous demands, but also based on their past allocations. To maintain past user allocation information, \name uses credits. 

Users earn credits in three ways. First, each user is bootstrapped with a fixed number of initial {\credit}s upon joining the system (we discuss the precise number once we have enough context, in \S\ref{ssec:algo-discussion}); second, each user is allocated $(1-\alpha) \cdot f$ free credits every quantum as compensation for contributing $(1-\alpha)$ fraction of its fair share to shared slices. Finally, users earn one credit when some other user borrows one of their {\em donated} slices (one credit per quantum per slice).

Unlike earning credits, there is only one way for any user to lose credits: for every slice borrowed from the {\mp} (donated or shared), the user loses one credit.

\vspace{-0.1in}
\subsubsection{Prioritized resource allocation}
\label{ssec:adr}
We now describe \name's resource allocation algorithm, that orchestrates resources and credits across users (Algorithm~\ref{algo:karma-algorithm}). To make the discussion succinct, we refer to the sum of user demands beyond their guaranteed share as ``borrower demand''; that is, to compute borrower demand for any given quantum, we take all users with demand greater than their guaranteed share and sum up the difference between their demand (in that quantum) and $\alpha \cdot f$. In quanta when borrower demand is equal to the supply (number of slices in {\mp}), \name's decision-making is trivial: simply allocate all slices in {\mp} to the borrowers, and update credits for all users as described in the previous subsection. The key algorithmic challenge that \name resolves is when the supply is either more or less than the borrower demand. We describe \name allocation mechanism for such scenarios next and then provide an illustrative example.

\begin{algorithm}[!t]
    \small
    \caption{
        {\bf: \name resource allocation algorithm.}
    }
    \label{algo:karma-algorithm}
    {\tt demand[u]}: demand of user \texttt{u} in the current quantum\\
    {\tt credits[u]}: credits of user \texttt{u}  in the current quantum\\
	{\tt alloc[u]}: allocation of user \texttt{u}  in the current quantum\\
    {$f$}: fair share\\
    {$\alpha$}: guaranteed fraction of fair share\\
    \\
    Every quantum do:
  \begin{algorithmic}[1]
  \State \texttt{shared\_slices} $\gets$ $n \cdot (1 - \alpha) \cdot f$
  \State For each user {\tt u}, 
  \State \hskip2em increment {\tt credits[u]} by $(1-\alpha)\cdot f$
  \State \hskip2em {\tt donated\_slices[u]} $=$ {\tt max} (0, $\alpha \cdot f -$ {\tt demand[u]})
  \State \hskip2em {\tt alloc[u]} $=$ {\tt min} ({\tt demand[u]}, {$\alpha \cdot f$})
  \State {\tt donors} $\gets$ all users {\tt u} with {\tt donated\_slices[u]} $> 0$\label{line:donors}
  \State {\tt borrowers} $\gets$ all users {\tt u} with\\\hskip3em {\tt alloc[u]}$<${\tt demand[u]} \& {\tt credits[u]}>0\label{line:borrowers}
  \vspace{0.1in}
  \While{{\tt borrowers} $\neq\phi$ and \\\hskip3em ($\sum_{u}$ {\tt donated\_slices[u]} > 0 or \texttt{shared\_slices} > 0)}\label{line:while-loop}
    \State $b^\star$ $\gets$ borrower with maximum {\tt credit}s
    \If{{\tt donors} $\neq \phi$}
      \State $d^\star$ $\gets$ donor with minimum {\tt credit}s
      \State Increment {\tt credits}[$d^\star$] by $1$ 
	  \State Decrement {\tt donated\_slices[u]} by $1$ 
	  \State Update the set of {\tt donors} (line $6$)\label{line:update-donors}
    \Else
      \State Decrement {\tt shared\_slices} by $1$
    \EndIf
    \State Increment {\tt alloc[$b^\star$]} by $1$
    \State Decrement {\tt credits}[$b^\star$] by $1$\label{line:decrement-credits}
    \State Update the set of {\tt borrowers} (line $7$)\label{line:update-borrowers}
   \EndWhile

\end{algorithmic}
\end{algorithm}

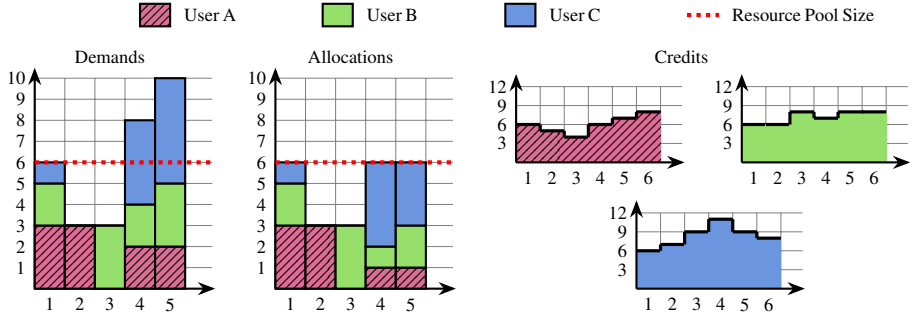
\begin{figure*}
  \centering
  \begin{tikzpicture}[xscale=0.4, yscale=0.28, font=\scriptsize, >=Stealth]

    \begin{scope}
    \draw[step=1,gray,very thin] (0, 0) grid (5.9, 10);
    \draw[->, thick] (0, 0) -- (6, 0);
    \draw[->, thick] (0, 0) -- (0, 10.5);
    \foreach \x/\xtext in {0.5/1, 1.5/2, 2.5/3, 3.5/4, 4.5/5}
        \node[below] at (\x, 0) {$\xtext$};
    \foreach \y in {1, 2,..., 10}
        \node[left] at (0, \y) {$\y$};
        
    \draw (0,0) pic[transform shape] {stackedbar={solidbox}{A}{B}{C}{3}{2}{1}};
    \draw[shift={(1,0)}] (0,0) pic[transform shape] {stackedbar={solidbox}{A}{B}{C}{3}{0}{0}};
    \draw[shift={(2,0)}] (0,0) pic[transform shape] {stackedbar={solidbox}{A}{B}{C}{0}{3}{0}};
    \draw[shift={(3,0)}] (0,0) pic[transform shape] {stackedbar={solidbox}{A}{B}{C}{2}{2}{4}};
    \draw[shift={(4,0)}] (0,0) pic[transform shape] {stackedbar={solidbox}{A}{B}{C}{2}{3}{5}};
    
    \draw[dotted, red, ultra thick] (0, 6) -- (6, 6); 
    
    \node at (2.5, 11) {Demands};
    \end{scope}
    
    \begin{scope}[shift={(8, 0)}]
    \draw[step=1,gray,very thin] (0, 0) grid (5.9, 10);
    \draw[->, thick] (0, 0) -- (6, 0);
    \draw[->, thick] (0, 0) -- (0, 10.5);
    \foreach \x/\xtext in {0.5/1, 1.5/2, 2.5/3, 3.5/4, 4.5/5}
        \node[below] at (\x, 0) {$\xtext$};
    \foreach \y in {1, 2,..., 10}
        \node[left] at (0, \y) {$\y$};
        
    \draw (0,0) pic[transform shape] {stackedbar={solidbox}{A}{B}{C}{3}{2}{1}};
    \draw[shift={(1,0)}] (0,0) pic[transform shape] {stackedbar={solidbox}{A}{B}{C}{3}{0}{0}};
    \draw[shift={(2,0)}] (0,0) pic[transform shape] {stackedbar={solidbox}{A}{B}{C}{0}{3}{0}};
    \draw[shift={(3,0)}] (0,0) pic[transform shape] {stackedbar={solidbox}{A}{B}{C}{1}{1}{4}};
    \draw[shift={(4,0)}] (0,0) pic[transform shape] {stackedbar={solidbox}{A}{B}{C}{1}{2}{3}};
    
    \draw[dotted, red, ultra thick] (0, 6) -- (6, 6); 
    
    \node at (2.5, 11) {Allocations};
    \end{scope}

    \begin{scope}[shift={(20, 0)}, yscale=0.3, xscale=0.8]
      \draw[ystep=3,xstep=1,gray,very thin] (0, 0) grid (6.9, 12);
      \foreach \x/\xtext in {0.5/1, 1.5/2, 2.5/3, 3.5/4, 4.5/5, 5.5/6}
          \node[below] at (\x, 0) {$\xtext$};
      \foreach \y in {3, 6,..., 12}
          \node[left] at (0, \y) {$\y$};
          
      \fill[C] (0, 0) rectangle (1, 6);
      \draw[black, very thick] (0, 6) -- +(1, 0);
      \fill[shift={(1,0)}, C] (0, 0) rectangle (1, 7);
      \draw[shift={(1,0)}, black, very thick] (0,7) -- +(1, 0) +(0, 0) -- +(0, -1);
      \fill[shift={(2,0)}, C] (0, 0) rectangle (1, 9);
      \draw[shift={(2,0)}, black, very thick] (0,9) -- +(1, 0) +(0, 0) -- +(0, -2);
      \fill[shift={(3,0)}, C] (0, 0) rectangle (1, 11);
      \draw[shift={(3,0)}, black, very thick] (0,11) -- +(1, 0) +(0, 0) -- +(0, -2);
      \fill[shift={(4,0)}, C] (0, 0) rectangle (1, 9);
      \draw[shift={(4,0)}, black, very thick] (0,9) -- +(1, 0) +(0, 0) -- +(0, +2);
      \fill[shift={(5,0)}, C] (0, 0) rectangle (1, 8);
      \draw[shift={(5,0)}, black, very thick] (0,8) -- +(1, 0) +(0, 0) -- +(0, +1);
      
      \draw[->, thick] (0, 0) -- (7, 0);
      \draw[->, thick] (0, 0) -- (0, 14);
    \end{scope}

    \begin{scope}[shift={(16, 6)}, yscale=0.3, xscale=0.8]
      \draw[ystep=3,xstep=1,gray,very thin] (0, 0) grid (6.9, 12);
      \foreach \x/\xtext in {0.5/1, 1.5/2, 2.5/3, 3.5/4, 4.5/5, 5.5/6}
          \node[below] at (\x, 0) {$\xtext$};
      \foreach \y in {3, 6,..., 12}
          \node[left] at (0, \y) {$\y$};
          
      \fill[A] (0, 0) rectangle (1, 6);
      \draw[black, very thick] (0, 6) -- +(1, 0);
      \fill[shift={(1,0)}, A] (0, 0) rectangle (1, 5);
      \draw[shift={(1,0)}, black, very thick] (0,5) -- +(1, 0) +(0, 0) -- +(0, +1);
      \fill[shift={(2,0)}, A] (0, 0) rectangle (1, 4);
      \draw[shift={(2,0)}, black, very thick] (0,4) -- +(1, 0) +(0, 0) -- +(0, +1);
      \fill[shift={(3,0)}, A] (0, 0) rectangle (1, 6);
      \draw[shift={(3,0)}, black, very thick] (0,6) -- +(1, 0) +(0, 0) -- +(0, -2);
      \fill[shift={(4,0)}, A] (0, 0) rectangle (1, 7);
      \draw[shift={(4,0)}, black, very thick] (0,7) -- +(1, 0) +(0, 0) -- +(0, -1);
      \fill[shift={(5,0)}, A] (0, 0) rectangle (1, 8);
      \draw[shift={(5,0)}, black, very thick] (0,8) -- +(1, 0) +(0, 0) -- +(0, -1);
      
      \draw[->, thick] (0, 0) -- (7, 0);
      \draw[->, thick] (0, 0) -- (0, 14);
    \end{scope}

    \begin{scope}[shift={(23.5, 6)}, yscale=0.3, xscale=0.8]
      \draw[ystep=3,xstep=1,gray,very thin] (0, 0) grid (6.9, 12);
      \foreach \x/\xtext in {0.5/1, 1.5/2, 2.5/3, 3.5/4, 4.5/5, 5.5/6}
          \node[below] at (\x, 0) {$\xtext$};
      \foreach \y in {3, 6,..., 12}
          \node[left] at (0, \y) {$\y$};
          
      \fill[B] (0, 0) rectangle (1, 6);
      \draw[black, very thick] (0, 6) -- +(1, 0);
      \fill[shift={(1,0)}, B] (0, 0) rectangle (1, 6);
      \draw[shift={(1,0)}, black, very thick] (0,6) -- +(1, 0);
      \fill[shift={(2,0)}, B] (0, 0) rectangle (1, 8);
      \draw[shift={(2,0)}, black, very thick] (0,8) -- +(1, 0) +(0, 0) -- +(0, -2);
      \fill[shift={(3,0)}, B] (0, 0) rectangle (1, 7);
      \draw[shift={(3,0)}, black, very thick] (0,7) -- +(1, 0) +(0, 0) -- +(0, +1);
      \fill[shift={(4,0)}, B] (0, 0) rectangle (1, 8);
      \draw[shift={(4,0)}, black, very thick] (0,8) -- +(1, 0) +(0, 0) -- +(0, -1);
      \fill[shift={(5,0)}, B] (0, 0) rectangle (1, 8);
      \draw[shift={(5,0)}, black, very thick] (0,8) -- +(1, 0) +(0, 0) -- +(0, 0);  

      \draw[->, thick] (0, 0) -- (7, 0);
      \draw[->, thick] (0, 0) -- (0, 14);
    \end{scope}

    \node at (21.5, 11) {Credits};
    
    \begin{scope}[shift={(4, 13)}, node distance=2]
    \node (legendA) [legenditem, A, label={[shift={(0.05,0)}]right:{User A}}] at (0, 0) {};
    \node (legendB) [legenditem, B, label={[shift={(0.05,0)}]right:{User B}}, right=of legendA] {};
    \node (legendC) [legenditem, C, label={[shift={(0.05,0)}]right:{User C}}, right=of legendB] {};
    \node (legendP) [legenditem, draw=white, minimum width=4mm, label={[shift={(0.05,0)}]right:{Resource Pool Size}}, right=of legendC] {};
    \draw[dotted, red, ultra thick] (legendP.west) -- (legendP.east);
    \end{scope}
    \end{tikzpicture}
  \vspace{-0.1in}
  \caption{\textbf{\name resource allocation for the running example of Figure~\ref{fig:max-min-example}:} Recall that there are $6$ resource slices, $3$ users each with average demand and fair share equal to $2$. We show the case of the guaranteed share being $1$ ($\alpha=0.5$), with $6$ bootstrapping (initial) credits for each user. Note that each user receives $1$ free credit every quantum. \name achieves significantly improved fair allocation than max-min fairness---it allocates each user an equal allocation of $8$ resource slices over time.} 
  \label{fig:fairness-example1-karma}
  \vspace{-0.1in}
  
\end{figure*}

\paragraphb{Orchestrating resources and credits when supply $>$ borrower demand} When supply is greater than borrower demand, there are enough slices in \mp to satisfy the demands of all borrowers. 
In such a case, \name prioritizes the allocation of donated slices over shared slices (so that donors get credits), and across multiple donated slices, prioritizes the allocation of a slice from the donor that has the smallest number of credits---this allows ``poorer'' donors to earn more credits, and moves the system towards a more balanced distribution of credits across users.
Intuitively, credits capture the allocation obtained by a user until the last quantum---users who obtained lower allocations in the past will have a higher than average (across users) number of credits, while those who received a surplus of allocations will have a below-average number of credits. Hence, balancing the number of credits across users over time allows \name to move towards a more equitable set of total allocations across users. 
Once all donated slices are allocated, \name allocates shared slices to satisfy the remaining borrower demands.

\paragraphb{Orchestrating resources and credits when supply $<$ borrower demand} When supply is less than demand, \mp does not have enough slices to satisfy all borrower demands. In such a scenario, \name prioritizes allocating slices to users with the maximum number of credits. This strategy essentially favors users that had fewer allocations in the past (and thus, a larger number of credits), hence moving the system towards a more balanced allocation of resources across users, promoting fairness. At the same time, reducing the credits for the users with the most credits also moves the system to a more balanced distribution of credits across users.

\paragraphb{Illustrative example} 
We now illustrate  through a concrete example. The running example in 
Figure~\ref{fig:fairness-example1-karma} shows the execution of \name's algorithm for the example from Figure~\ref{fig:max-min-example} for $\alpha=0.5$: that is three users A, B, and C, each with a fair share $2$ slices ($f=2$), and a guaranteed share of $1$ slice. Recall that, since $(1-\alpha)\cdot f = 1$, each user receives $1$ credit every quantum, and suppose all users are bootstrapped with $6$ initial credits.

In the first quantum, C's demand is equal to the guaranteed share, while A and B request $2$ and $1$ slices beyond the guaranteed share, respectively. Since supply ($=3$ shared slices in {\mp}) is equal to borrower demand, \name uses the shared slices to allocate slices beyond the guaranteed share for A and B and satisfies their demands. This results in a final allocation of $3$ slices for $A$, $2$ slices for $B$, and $1$ slice for $C$. $A$ loses $2$ credits, and $B$ loses $1$ credits, and no one gains any credits.

In the second quantum, A demands $3$ slices, while B and C donate $1$ slice each. The total supply ($=5$, with $2$ donated slices and $3$ shared slices) exceeds the borrower demand. A is allocated $3$ slices and it loses $3$ credits (since its allocation is $2$ slices above its guaranteed share). B and C receive $1$ credit each since their donated slices are used. Similarly, in the third quantum, B demands $3$ slices, while A and C donate $1$ slice each. Since total supply exceeds borrower demand, B receives the $3$ slices it asked for, and loses $2$ credits; A and C gain $1$ credit each. 

The fourth quantum is important: here, demand exceeds supply, and there are no donated slices. Now, unlike classic max-min fairness, \name will prioritize the allocation of resources based on the credits of each tenant. Since at the start of this quantum, C has $11$ credits, while A and B have only $6$ and $7$ credits respectively, C will be able to get $3$ extra slices from the pool of shared slices by using $3$ credits and achieve an allocation of $4$. A and B will get their guaranteed allocation of $1$ and do not gain or lose any credits.

In the fifth quantum, once again, demand exceeds supply. C has $9$ credits, B has $8$ credits, and A has $7$ credits. \name first prioritizes allocating to C giving it $1$ extra slice, at which point both C and B have equal credits ($8$). Next, they both get $1$ extra slice each, at which point the supply is exhausted. The final resulting allocation is $1$ slice for A, $2$ slices for B, and $3$ slices for C.

In the end, A, B, and C end up with the exact same total allocation ($8$ slices) and number of credits (unlike max-min fairness where user allocations had a disparity of $2\times$).

  \vspace{-0.1in}
\subsection{\name Properties \& Guarantees}
\label{ssec:formal}
  \vspace{-0.05in}

\begin{figure*}
  \centering
  \begin{tikzpicture}[x=0.4cm, y=0.16cm, font=\scriptsize, >=Stealth]

    \begin{scope}
        \begin{scope}
        \draw[step=2,gray,very thin] (0, 0) grid (3.9, 16);
        \draw[->, thick] (0, 0) -- (4, 0);
        \draw[->, thick] (0, 0) -- (0, 17);
        \foreach \x/\xtext in {0.5/1, 1.5/2, 2.5/3}
            \node[below] at (\x, 0) {$\xtext$};
        \foreach \y in {2, 4,..., 16}
            \node[left] at (0, \y) {$\y$};
            
        \draw (0,0) pic[transform shape] {stackedbar4={solidbox}{A}{B}{C}{D}{8}{8}{0}{0}};
        \draw[shift={(1,0)}] (0,0) pic[transform shape] {stackedbar4={solidbox}{A}{B}{C}{D}{8}{0}{8}{0}};
        \draw[shift={(2,0)}] (0,0) pic[transform shape] {stackedbar4={solidbox}{A}{B}{C}{D}{8}{8}{0}{0}};
        
        \draw[dotted, red, ultra thick] (0, 8) -- (4, 8); 
        
        \node at (1.5, 18) {Demands};
        \end{scope}
        
        \begin{scope}[shift={(8, 9.5)}, yscale=0.85]
        \draw[step=2,gray,very thin] (0, 0) grid (3.9, 7.9);
        \draw[->, thick] (0, 0) -- (4, 0);
        \draw[->, thick] (0, 0) -- (0, 9);
        \foreach \x/\xtext in {0.5/1, 1.5/2, 2.5/3}
            \node[below] at (\x, 0) {$\xtext$};
        \foreach \y in {2, 4,..., 8}
            \node[left] at (0, \y) {$\y$};
            
        \draw (0,0) pic[transform shape] {stackedbar4={solidbox}{A}{B}{C}{D}{4}{4}{0}{0}};
        \draw[shift={(1,0)}] (0,0) pic[transform shape] {stackedbar4={solidbox}{A}{B}{C}{D}{2}{0}{6}{0}};
        \draw[shift={(2,0)}] (0,0) pic[transform shape] {stackedbar4={solidbox}{A}{B}{C}{D}{3}{5}{0}{0}};
        
        \draw[dotted, red, ultra thick] (0, 8) -- (4, 8); 
        
        \node at (1.5, 10) {Allocations};
        \end{scope}
        
        \begin{scope}[shift={(8, -0.5)}, yscale=0.85]
        \draw[step=2,gray,very thin] (0, 0) grid (3.9, 7.9);
        \draw[->, thick] (0, 0) -- (4, 0);
        \draw[->, thick] (0, 0) -- (0, 9);
        \foreach \x/\xtext in {0.5/1, 1.5/2, 2.5/3}
            \node[below] at (\x, 0) {$\xtext$};
        \foreach \y in {2, 4,..., 8}
            \node[left] at (0, \y) {$\y$};
            
        \draw (0,0) pic[transform shape] {stackedbar4={solidbox}{A}{B}{C}{D}{0}{8}{0}{0}};
        \draw[shift={(1,0)}] (0,0) pic[transform shape] {stackedbar4={solidbox}{A}{B}{C}{D}{4}{0}{4}{0}};
        \draw[shift={(2,0)}] (0,0) pic[transform shape] {stackedbar4={solidbox}{A}{B}{C}{D}{6}{2}{0}{0}};
        
        \draw[dotted, red, ultra thick] (0, 8) -- (4, 8); 
        
        \end{scope}
        
        \draw[->, very thick] (4,8) -- node[above, sloped] {A is honest} +(60:6.5);
        \draw[->, very thick] (4,8) -- node[below, sloped] {A under-reports} +(-60:6.5);
    \end{scope}
    
    \begin{scope}[shift={(20,0)}]
        \begin{scope}
        \draw[step=2,gray,very thin] (0, 0) grid (3.9, 16);
        \draw[->, thick] (0, 0) -- (4, 0);
        \draw[->, thick] (0, 0) -- (0, 17);
        \foreach \x/\xtext in {0.5/1, 1.5/2, 2.5/3}
            \node[below] at (\x, 0) {$\xtext$};
        \foreach \y in {2, 4,..., 16}
            \node[left] at (0, \y) {$\y$};
            
        \draw (0,0) pic[transform shape] {stackedbar4={solidbox}{A}{B}{C}{D}{8}{0}{0}{0}};
        \draw[shift={(1,0)}] (0,0) pic[transform shape] {stackedbar4={solidbox}{A}{B}{C}{D}{8}{2}{2}{2}};
        \draw[shift={(2,0)}] (0,0) pic[transform shape] {stackedbar4={solidbox}{A}{B}{C}{D}{8}{2}{2}{2}};
        
        \draw[dotted, red, ultra thick] (0, 8) -- (4, 8); 
      
        \node at (1.5, 18) {Demands};
        \end{scope}
        
        \begin{scope}[shift={(8, 9.5)}, yscale=0.85]
        \draw[step=2,gray,very thin] (0, 0) grid (3.9, 7.9);
        \draw[->, thick] (0, 0) -- (4, 0);
        \draw[->, thick] (0, 0) -- (0, 9);
        \foreach \x/\xtext in {0.5/1, 1.5/2, 2.5/3}
            \node[below] at (\x, 0) {$\xtext$};
        \foreach \y in {2, 4,..., 8}
            \node[left] at (0, \y) {$\y$};
            
        \draw (0,0) pic[transform shape] {stackedbar4={solidbox}{A}{B}{C}{D}{8}{0}{0}{0}};
        \draw[shift={(1,0)}] (0,0) pic[transform shape] {stackedbar4={solidbox}{A}{B}{C}{D}{2}{2}{2}{2}};
        \draw[shift={(2,0)}] (0,0) pic[transform shape] {stackedbar4={solidbox}{A}{B}{C}{D}{2}{2}{2}{2}};
        
        \draw[dotted, red, ultra thick] (0, 8) -- (4, 8); 
        
        \node at (1.5, 10) {Allocations};
        \end{scope}
        
        \begin{scope}[shift={(8, -0.5)}, yscale=0.85]
        \draw[step=2,gray,very thin] (0, 0) grid (3.9, 7.9);
        \draw[->, thick] (0, 0) -- (4, 0);
        \draw[->, thick] (0, 0) -- (0, 9);
        \foreach \x/\xtext in {0.5/1, 1.5/2, 2.5/3}
            \node[below] at (\x, 0) {$\xtext$};
        \foreach \y in {2, 4,..., 8}
            \node[left] at (0, \y) {$\y$};
            
        \draw (0,0) pic[transform shape] {stackedbar4={solidbox}{A}{B}{C}{D}{0}{0}{0}{0}};
        \draw[shift={(1,0)}] (0,0) pic[transform shape] {stackedbar4={solidbox}{A}{B}{C}{D}{2}{2}{2}{2}};
        \draw[shift={(2,0)}] (0,0) pic[transform shape] {stackedbar4={solidbox}{A}{B}{C}{D}{2}{2}{2}{2}};
        
        \draw[dotted, red, ultra thick] (0, 8) -- (4, 8); 
        
        \end{scope}
    
        \draw[->, very thick] (4,8) -- node[above, sloped] {A is honest} +(60:6.5);
        \draw[->, very thick] (4,8) -- node[below, sloped] {A under-reports} +(-60:6.5);
    \end{scope}

    \begin{scope}[shift={(5, 22)}, node distance=3]
      \node (legendA) [legenditem, A, label={[shift={(0.05,0)}]right:{User A}}] at (0, 0) {};
      \node (legendB) [legenditem, B, label={[shift={(0.05,0)}]right:{User B}}, right=of legendA] {};
      \node (legendC) [legenditem, C, label={[shift={(0.05,0)}]right:{User C}}, right=of legendB] {};
      \node (legendD) [legenditem, D, label={[shift={(0.05,0)}]right:{User D}}, right=of legendC] {};
      \node (legendP) [legenditem, draw=white, minimum width=4mm, label={[shift={(0.05,0)}]right:{Resource Pool Size}}, right=of legendD] {};
      \draw[dotted, red, ultra thick] (legendP.west) -- (legendP.east);
      \end{scope}

    \end{tikzpicture}
  \vspace{-0.1in}
  \caption{\textbf{The phenomenon of users (left) gaining a small factor of improvement in their allocations by specifying demands less than their real demands, by exploiting knowledge of all future demands of all users; (right) any imprecision in the knowledge of future demands of all users could result in a significant reduction in useful allocations of the lying user.} The resource pool has $8$ slices, and $4$ users with fair share of $2$ and guaranteed share of 0 ($\alpha=0$). %
  } 
  \label{fig:under-reporting-example}
  \vspace{-0.1in}
\end{figure*}
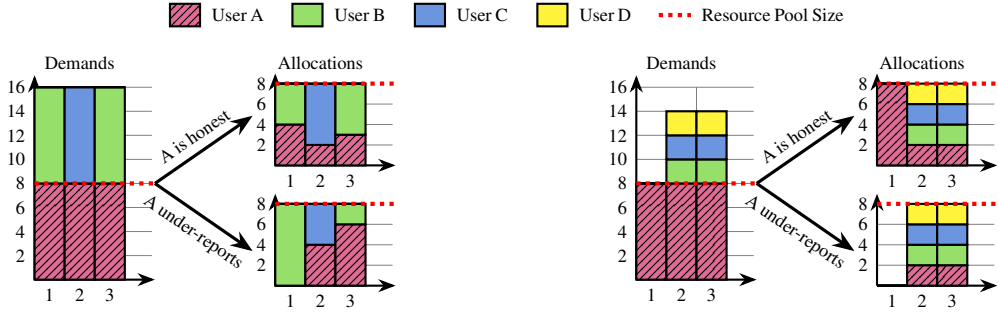

In this section, we present a theoretical analysis of \name. Recall from \S\ref{ssec:mempool} that, similar to all prior works, users are considered selfish and strategic (that is, are willing to misreport their demands to maximize their allocations), but not adversarial (that is, do not lie about their demands simply to hurt others' allocations). For the purpose of our theoretical analysis, we assume that \name is initialized with a large enough number of initial credits so that users do not run out of credits during the execution of the algorithm (we discuss how to achieve this in practice in \S~\ref{ssec:algo-discussion}). All our results hold for $\alpha=0$; extending our results to $\alpha>0$ is an interesting open question. Finally, while we provide inline intuition for each of our results, full proofs are presented in~\S\ref{app:proofs}.

We define Pareto efficiency on a per-quantum basis. An allocation is said to be Pareto efficient if it is not possible to increase the allocation of a user without decreasing the allocation of at least one other user by a similar total amount during that quantum. Note that, Pareto efficiency on a per-quantum basis implies Pareto efficiency over time. 
  
\begin{theorem}\label{thm:pareto}
  \name is Pareto efficient.
\end{theorem}

\noindent
\name's Pareto efficiency follows trivially from the observation that similar to max-min fairness, \name allocation satisfies the two properties: (1) no user is allocated more resources than its demand, and (2) either all resources are allocated or all demands are satisfied.

For strategy-proofness, we make two important notes. First, if one assumes that the system has a priori knowledge of all future user demands, the resource allocation problem can be solved trivially using dynamic programming; however, for many use cases, it is hard to have a priori knowledge of all future user demands. This leads to our second note: \name is solving an ``online'' problem (that is, it does not assume a priori knowledge of future user demands), and thus, we prove online strategy-proofness~\cite{aleksandrov2019strategy} defined as follows: assume that all users are honest during quanta $0$ to $q-1$; then, a mechanism is said to be online strategy-proof if, for any quantum $q$, a user cannot increase its allocation during quantum $q$ by lying about its demand during quantum $q$.

\begin{theorem}\label{thm:strategy-proof}
  \name is online strategy-proof.
\end{theorem}

\noindent
To prove Theorem~\ref{thm:strategy-proof}, we actually prove a stronger result stated below. \name's online strategy-proofness trivially follows from this.

\begin{lemma}\label{thm:1}
  A user cannot increase its useful resource allocation by specifying a demand higher than its real demand in any quantum. %
\end{lemma}

\noindent
The proof for the lemma is a bit involved, but intuitively, it shows the following. The immediate effect of a user specifying a demand higher than its actual demand is that if the user is allocated more resources than its actual demand, these extra resources do not contribute to its utility, but do put the user into a disadvantageous position: not only can this user lose credits (either because it's asking for resources beyond its guaranteed share, or because it could have gained credits if this extra resource could have been allocated to some borrower), but also because other users get fewer resources; this makes other users be favored by the allocation algorithm in the future while making the lying user less favored. Thus, the user cannot increase its long-term ``useful'' allocation by specifying a demand higher than the real demand in any quantum. Specifically, it is possible that when a user over-reports its demand during quantum $q'$, the user receives an increased instantaneous allocation during some future quantum $q > q'$; however, we are able to show that, in this case, the user will also receive reduced instantaneous allocation(s) during other quantum(s) in between $q'$ and $q$, leading to either a lower or equal total allocation over the period between $q'$ and $q$. The hardness in the proof stems from carefully analyzing such cascade effects: a small change in users' resource allocation in any quantum can result in complex changes in future allocations that may lead to higher instantaneous but equal or lower total allocations in future quanta. Once we prove this lemma, the proof for \name's online strategy-proofness follows immediately.

While analyzing \name properties, we encountered a new, surprising, phenomenon that may be of further theoretical interest: we show that a user that {\em knows all future demands of all other users} can report a demand that is lower than its actual demand in the current quantum to increase its allocation in future quanta by a small constant factor. However, any imprecision in the knowledge of all future demands of all other users could result in the user losing a factor of $\Omega(n)$ of its total allocation.

\begin{lemma}\label{thm:more_than_triple}
   A user cannot increase its total useful allocation by a factor more than $1.5\times$ by specifying a demand less than its real demand in any quantum. Gaining this useful allocation requires the user to know the future demands of all users. If the user does not have a precise knowledge of all future demands of all users, it can lose its useful allocation by a factor of $\frac{n+2}{2}$ (for $n \geq 3$) by specifying a demand less than its real demand.
\end{lemma}

\noindent
We provide intuition for this phenomenon using an example (Figure~\ref{fig:under-reporting-example}). In the left figure, user A is able to gain 1 extra slice in its overall allocation by under-reporting its demand (reporting $0$ instead of $8$) in the first quantum. By under-reporting, its allocation in the first quantum reduces, enabling it to get more resources during the second quantum when it competes with user C. In the third quantum, it is able to recover the resources it lost in the first quantum from user B, resulting in an overall gain. To see the flip-side, if the demands of other users had been as shown in Figure~\ref{fig:under-reporting-example}~(right), then user A sees a $3\times$ degradation in overall allocation.

To prove the first part of the Lemma~\ref{thm:more_than_triple}, we consider an arbitrary user Alice and an arbitrary time period, and compare two scenarios---one where Alice is truthful (hereby called the truthful scenario) and one where Alice is deviating by under-reporting her demand during some quantum (hereby called the deviating scenario).

Our key insight for the proof is that bounding the increase in total allocation of \textit{all users} is easier than reasoning about the increase in total allocation of an individual user (Alice) since even a small change in Alice's demand during one quantum can result in cascading effects on the total allocation of other users as well. To that end, we prove the following claim: the total amount of resources all the users have earned in excess in the deviating scenario compared to the truthful one can be at most as large as Alice's total allocation in the truthful scenario. We prove this claim based on the following observation: whenever Alice under-reports her demand she is effectively "donating" the allocation she would have gotten in the truthful scenario to the other users whose allocations in the deviating scenario increase. Since \name is Pareto efficient, the total gain in allocation across users during this quantum is limited by the amount donated by Alice which is in turn bound by Alice's own allocation during this quantum in the truthful scenario. By applying this reasoning iteratively across all quanta\footnote{It turns out that Alice under-reporting in a given quantum cannot cause cascading increases in total allocation across users in future quanta if Alice does not under-report in future quanta. This is because \name prioritizes allocation to users with high credits (or equivalently low total allocations).}, we can show that the total increase in allocation across all users cannot exceed the total allocation of Alice in the truthful scenario. This already implies a $2\times$ upper bound on the maximum increase in total allocation that Alice can achieve.

To tighten the upper bound, we prove a second claim: if Alice receives higher total allocation in the deviating scenario compared to the truthful scenario, then there must exist some other user Bob who gained an even larger increase in total allocation than Alice. 
Putting together the above two claims allows us to establish the desired upper bound. Based on the first claim, the total gain in allocation across all users cannot exceed Alice's total allocation in the truthful scenario. This implies that the sum of total gains across Alice and Bob cannot exceed Alice's total allocation in the truthful scenario. Since Bob's gain is at least as large as Alice's gain (based on the second claim), this implies that Alice's gain is at most half the total allocation of Alice in the truthful scenario---a gain of at most $1.5\times$, thus proving the first part of Lemma~\ref{thm:more_than_triple}.

The second part of the lemma is proven by first creating a set of demands where a user can under-report its demand during quantum $q$ to earn increased total allocation by some quantum $q' > q$. Then we create a set of demands that are identical up to quantum $q$ but vastly different from quanta $q+1$ to $q'$. If the user (in the hope of facing the first set of demands) under-reports its demand on quantum $q$ but ends up facing the second set of demands then this results in vastly different allocations by quantum $q'$. By correctly picking the two sets of demands we get the desired bounds.

In \S\ref{app:proofs}, we prove an even stronger result that extends \name properties from Theorem~\ref{thm:pareto}, Theorem~\ref{thm:strategy-proof}, Lemma~\ref{thm:1} and Lemma~\ref{thm:more_than_triple} to the case of multiple colluding users:

\begin{theorem}\label{thm:collusion}
  No group of colluding users can increase their allocation by specifying a demand higher than their real demand. Additionally, for any group of colluding users, under-reporting demands cannot lead to more than a $2\times$ improvement in their useful resource allocation. Finally, even if users form coalitions, \name is Pareto efficient and online strategy-proof.
\end{theorem}

\noindent
Recall that \name focuses on long-term fairness without a priori knowledge of future user demands. To that end, the following theorem summarizes \name's fairness guarantees:

\begin{theorem}\label{thm:karma_lt_fariness}
  For any quantum $q$, given fixed user allocations from quantum $0$ to quantum $q-1$, and user demands at quantum $q$, Karma maximizes the minimum total allocation from quantum $0$ to quantum $q$ across users.
\end{theorem}

The proof for the above theorem follows from the prioritized resource allocation mechanism of \name. Intuitively, given allocations from quantum $0$ to $q-1$, the user with the least total allocation up to quantum $q-1$ will have the largest number of credits. In quantum $q$, \name will prioritize the allocation of resources to this user (until it is no longer the one with the minimum total allocation, after which it will prioritize the next user with the minimum total allocation, and so on), thus maximizing the minimum total allocation from quantum $0$ to $q$ across users---this is the best one can do in quantum $q$ given past allocations.

\subsection{Discussion}
\label{ssec:algo-discussion}
Finally, we briefly discuss some additional aspects of \name design not included in the previous subsections. 

\paragraphb{Bootstrapping \name with initial credits}
Recall that, to bootstrap users, \name allocates each user an initial number of credits. The precise number of initial credits has little impact on {\name}'s behavior; after all, credits in \name essentially capture a relative ordering between users, rather than having any absolute meaning. The only importance of the number of credits is to ensure that no user runs out of credits at any quantum (which, in turn, could lead to violation of {\name}'s Pareto efficiency guarantees): even if spare resources are available, a user with high demand may not be able to borrow resources beyond the guaranteed share (line 7 of Algorithm~\ref{algo:karma-algorithm}) due to running out of credits. Thus, \name sets the number of initial credits to a large numerical value to ensure that no user ever runs out of credits\footnote{For example, in a system with $100$ users with fair share of $100$ slices, setting initial credits to say $10^{13}$ will ensure that even a worst-case user with highest possible demand ($10000$ slices) during all quanta cannot run out of credits for $\sim31$ years, which is good enough for all practical purposes.},

\paragraphb{User churn}
Fairness is relatively ill-defined when users can join and leave the system on a short-term basis (\eg, when a user runs a short query with large parallelism, and then leaves the cluster). Also, recall from our motivating scenarios, fair resource allocation in private clouds is usually performed for long-running services. However, {\name} still handles user churn since, in many realistic scenarios, the set of all users of the system may not be known upfront during system initialization. For users that join and leave over longer timescales, \name handles user churn with a simple mechanism: its credits. When a new user joins, either the resource pool size remains fixed and the fair share of all users is reduced proportionally or the resource pool size increases and the fair share of users remains the same. The credits of the existing $n-1$ users do not change, and the new user is bootstrapped with initial credits equal to the current average number of credits across the existing $n-1$ users. Intuitively, users who have donated more resources than they have borrowed will have above-average credits, and those who have borrowed more than they have donated will have below-average credits. As such, initializing the new user with the average number of credits (heuristically) puts the new user on equal footing with an existing user that has borrowed and donated equal amounts of resources over time. When a user leaves the system, the fair share of the remaining users is increased proportionally (or resource pool size reduces while maintaining the same fair share), and there is no change in their credits.

\paragraphb{Users with different fair shares}
We have presented \name's algorithm for the case of users having the same fair share merely for simplicity: all our results extend to the case of users having different fair shares. To generalize the algorithm to users with different fair shares, users with larger weights are charged fewer credits to borrow resources beyond their guaranteed share when compared to users with smaller weights. Intuitively, this enables users with larger weights to obtain more resources than users with smaller weights for the same number of credits. We achieve this by updating Line~\ref{line:decrement-credits} of Algorithm~\ref{algo:karma-algorithm} to decrement credits by $\frac{1}{n\cdot w_i}$ instead of $1$, where $w_i$ is the normalized weight of the corresponding user, and $n$ is the number of users. For users with different fair shares, this generalization leads to the same properties and guarantees as discussed in \S\ref{ssec:formal} (the only difference, is that the upper bound factor in Lemma~\ref{thm:more_than_triple} changes from $1.5\times$ to $2\times$). A full description of the weighted version of the algorithm along with proofs of guarantees can be found in \S\ref{appendix:weighted-algorithm}.

\paragraphb{System parameters, and interpretation for $\alpha$}
\name has only one parameter: $\alpha$; one can think of resource slice size and quantum duration as parameters, but these are irrelevant to \name's guarantees: they hold for any slice size and quantum duration, as long as demands change at coarse timescales than the quantum duration. 
The $\alpha$ parameter in \name provides a tradeoff between instantaneous and long-term fairness. Providers can choose any $\alpha$ depending on the desired properties. Intuitively, an $\alpha$ smaller than $1$ leads to a larger portion of shared slices, giving \name's algorithm more flexibility in adjusting allocations to achieve better long-term fairness. 

\section{\name Implementation Details}
\label{sec:arch}
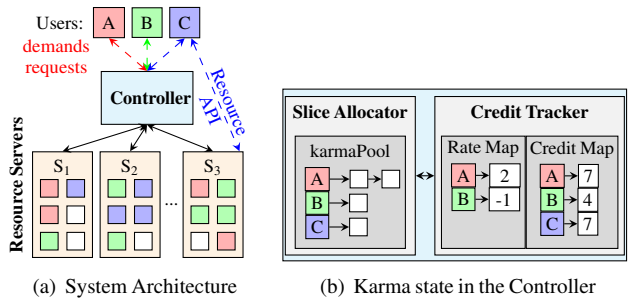
\begin{figure}
  \subfigure[System Architecture] {
  \begin{tikzpicture}[font=\scriptsize]
    \node[draw, fill=cyan!10, text width=2.8em, minimum height=2em] (cp) {\textbf{Controller}};
    
    \node[draw, fill=BurntOrange!10, text width=1.6em, minimum height=4em, below=1em of cp, xshift=-3.2em] (s1) {};
    \node[draw, fill=BurntOrange!10, text width=1.6em, minimum height=4em, right=0.25em of s1] (s2) {};
    \node[right=-0.2em of s2] (dots) {...};
    \node[draw, fill=BurntOrange!10, text width=1.6em, minimum height=4em, right=-0.2em of dots] (s3) {};
    
    \node[left=0em of s1, yshift=0.75em] {\rotatebox{90}{\textbf{Resource Servers}}};
    
    \node[draw, fill=red!30, below=1.05em of s1.north, xshift=-.5em] {};
    \node[draw, fill=blue!30, below=1.05em of s1.north, xshift=.5em] {};
    \node[draw, fill=red!30, below=2.05em of s1.north, xshift=-.5em] {};
    \node[draw, fill=white!30, below=2.05em of s1.north, xshift=.5em] {};
    \node[draw, fill=green!30, below=3.05em of s1.north, xshift=-.5em] {};
    \node[draw, fill=white!30, below=3.05em of s1.north, xshift=.5em] {};
    \node[below=-0.05em of s1.north] {$\text{S}_1$};
    
    \node[draw, fill=green!30, below=1.05em of s2.north, xshift=-.5em] {};
    \node[draw, fill=blue!30, below=1.05em of s2.north, xshift=.5em] {};
    \node[draw, fill=blue!30, below=2.05em of s2.north, xshift=-.5em] {};
    \node[draw, fill=blue!30, below=2.05em of s2.north, xshift=.5em] {};
    \node[draw, fill=green!30, below=3.05em of s2.north, xshift=-.5em] {};
    \node[draw, fill=white!30, below=3.05em of s2.north, xshift=.5em] {};
    \node[below=-0.05em of s2.north] {$\text{S}_2$};
    
    \node[draw, fill=red!30, below=1.05em of s3.north, xshift=-.5em] {};
    \node[draw, fill=green!30, below=1.05em of s3.north, xshift=.5em] {};
    \node[draw, fill=green!30, below=2.05em of s3.north, xshift=-.5em] {};
    \node[draw, fill=green!30, below=2.05em of s3.north, xshift=.5em] {};
    \node[draw, fill=white!30, below=3.05em of s3.north, xshift=-.5em] {};
    \node[draw, fill=red!30, below=3.05em of s3.north, xshift=.5em] {};
    \node[below=-0.05em of s3.north] {$\text{S}_3$};
    
    \draw[stealth-stealth] (cp.south) -- (s1.north);
    \draw[stealth-stealth] (cp.south) -- (s2.north);
    \draw[stealth-stealth] (cp.south) -- (s3.north);
    
    \node[draw, fill=green!30, above=1.25em of cp] (t2) {B};
    \node[draw, fill=red!30, left=0.25em of t2] (t1) {A};
    \node[draw, fill=blue!30, right=0.25em of t2] (t3) {C};
    
    \node[left=0em of t1] {Users:};
    
    \draw[red, dashed, stealth-stealth] (cp.north) -- node [pos=0.4, left=1em, text width=4.5em, align=center, text width=3em] {demands\\requests} (t1.south);
    \draw[green, dashed, stealth-stealth] (cp.north) -- (t2.south);
    \draw[blue, dashed, stealth-stealth] (cp.north) -- (t3.south);
    \draw[blue, dashed, stealth-stealth] ($(t3.south)+(0.25em, 0em)$) -- node[pos=0.2, right=0em, text width=3em, align=center, sloped] {Resource\\API} ($(s3.north)+(1em, 0em)$);    
  \end{tikzpicture}
  \label{fig:arch}
  }%
  \subfigure[\name state in the Controller] {
  \begin{tikzpicture}[font=\scriptsize]
    \node[draw, fill=cyan!10, minimum height=6.5em, minimum width=13em] (cp-detail) {};
    
    \node[draw, fill=gray!10, text width=4.25em, minimum height=6em, below=0.28em of cp-detail.north, xshift=-3.95em] (as) {};
    \node[below=0em of as.north] (as-title) {\textbf{Slice Allocator}};
    \node[draw, fill=gray!30, text width=3.5em, minimum height=4.25em, below=1.5em of as.north] (mp) {};
    \node[below=0em of mp.north] (mp-title) {\mp};
    \node[draw, inner sep=2pt, fill=red!30, below=0em of mp-title.south, xshift=-1.25em] (mp1) {A};
    \node[draw, inner sep=2pt, fill=green!30, below=0em of mp1] (mp2) {B};
    \node[draw, inner sep=2pt, fill=blue!30, below=0em of mp2] (mp3) {C};
    \node[draw, fill=white, right=0.75em of mp1] (mpb11) {};
    \node[draw, fill=white, right=0.5em of mpb11] (mpb12) {};
    \node[draw, fill=white, right=0.75em of mp2] (mpb21) {};
    \node[draw, fill=white, right=0.75em of mp3] (mpb31) {};
    
    \draw[-stealth] (mp1.east) -- (mpb11.west);
    \draw[-stealth] (mpb11.east) -- (mpb12.west);
    \draw[-stealth] (mp2.east) -- (mpb21.west);
    \draw[-stealth] (mp3.east) -- (mpb31.west);
    
    \node[draw, fill=gray!10, text width=6.5em, minimum height=6em, right=0.7em of as] (ut) {};
    \node[below=0em of ut.north] (ut-title) {\textbf{Credit Tracker}};
    \node[draw, fill=gray!30, text width=2.5em, minimum height=4.25em, below=1.5em of ut.north, xshift=-1.75em] (rm) {};
    \node[inner sep=2pt, below=0em of rm.north] (rm-title) {Rate Map};
    \node[draw, inner sep=2pt, fill=red!30, below=0em of rm-title.south, xshift=-0.75em] (t1) {A};
    \node[draw, inner sep=2pt, fill=green!30, below=0em of t1] (t2) {B};
    \node[draw, inner sep=2pt, align=center, text width=0.75em, fill=white, right=0.5em of t1] (u1) {2};
    \node[draw, inner sep=2pt, align=center, text width=0.75em, fill=white, right=0.5em of t2] (u2) {-1};
    \draw[-stealth] (t1.east) -- (u1.west);
    \draw[-stealth] (t2.east) -- (u2.west);
    
    \node[draw, fill=gray!30, text width=2.8em, minimum height=4.25em, right=0em of rm] (cm) {};
    \node[inner sep=2pt, below=0em of cm.north] (cm-title) {{Credit Map}};
    \node[draw, inner sep=2pt, fill=red!30, below=0em of cm-title.south, xshift=-0.75em] (ct1) {A};
    \node[draw, inner sep=2pt, fill=green!30, below=0em of ct1] (ct2) {B};
    \node[draw, inner sep=2pt, fill=blue!30, below=0em of ct2] (ct3) {C};
    \node[draw, inner sep=2pt, align=center, fill=white, right=0.5em of ct1] (c1) {7};
    \node[draw, inner sep=2pt, align=center, fill=white, right=0.5em of ct2] (c2) {4};
    \node[draw, inner sep=2pt, align=center, fill=white, right=0.5em of ct3] (c3) {7};
    \draw[-stealth] (ct1.east) -- (c1.west);
    \draw[-stealth] (ct2.east) -- (c2.west);
    \draw[-stealth] (ct3.east) -- (c3.west);

    \draw[stealth-stealth] (as.east) -- (ut.west);
  \end{tikzpicture}
  \label{fig:controller}
  }
  \vspace{-0.1in}
  \caption{\textbf{\name Design.} See \S\ref{sec:arch} for details.} 
  \vspace{-0.1in}
\end{figure}

We have implemented \name on top of Jiffy~\cite{jiffy}, an open-sourced elastic far memory system. Jiffy has a standard distributed data store architecture (Figure~\ref{fig:arch}): resources are partitioned into fixed-sized slices (blocks of memory) across a number of resource servers (memory servers), identified by their unique sliceIDs (referred to as blockIDs in Jiffy). A logically centralized controller tracks the available and allocated slices across the various resource servers and stores a mapping that translates sliceIDs to the corresponding resource server. We have implemented \name as a new resource allocation algorithm at the Jiffy controller\footnote{\name can thus directly piggyback on Jiffy's existing mechanisms for controller fault tolerance~\cite[Section 4]{jiffy} to persist its state across failures.}.

Users interact with the system through a client library that provides APIs for requesting resource allocation and accessing allocated resource slices. Users express their demands to the controller through resource requests which specify the number of slices required. The controller periodically performs resource allocation using the \name algorithm and provides users with the sliceIDs of the resource slices that are allocated to them. Users can then directly access these slices from the resource servers through read or write API calls without requiring controller interposition. In the rest of this section, we discuss the key data structures and mechanisms required to integrate \name with Jiffy.

\name employs three key data structures to efficiently implement the policies and mechanisms outlined in \S\ref{ssec:altruism}: \mp, a credit map, and a rate map.

\paragraphb{\mp} Recall from \S\ref{ssec:incentive} that the \mp tracks the pool of donated slices and shared slices, and needs to be updated when resource allocations change. Also, the resource allocation algorithm should be able to efficiently select donated slices from a particular user while satisfying borrower demands (\S\ref{ssec:adr}). To this end, the \mp is implemented as a hash map, mapping userIDs to the list of sliceIDs corresponding to slices donated by them. The list of sliceIDs corresponding to shared slices is stored in a separate entry of the same hash map. When resource allocations change, the corresponding sliceIDs are added to or removed from the corresponding lists. As such, \mp supports all updates in $O(1)$ time.

\paragraphb{Credit Tracking}
\name employs two data structures for tracking and allocating credits across various users: a rate map and a credit map. The rate map maps each user to the \textit{rate} at which it earns or spends its credits every quantum, that is, the difference between the user's guaranteed share and the number of its allocated slices in that quantum. The rate is positive when the user is earning (that is, has donated slices) and negative when it is spending credits (that is, has borrowed slices), respectively. The credit map, on the other hand, maps each user to a counter corresponding to its current credits.

Separating the rate map and credit map facilitates efficient credit tracking at each quantum: \name simply iterates through the rate map entries, and updates the credit counters in the credit map based on the corresponding user credit rates. Since the rate map only contains entries for users with non-zero rates, \name can efficiently update credits for only the relevant users. At the same time, Employing a hash-map for each of them permits $O(1)$ updates to the user credit rate or number of credits while performing resource allocation.

\paragraphb{Borrowing and donating slices} \name realizes its credit-based prioritized allocation algorithm (\S\ref{ssec:incentive}) using two modules at the controller. First is a \textit{slice allocator} that maintains the \mp to track and update slice allocations across users, and, second a \textit{credit tracker} that maintains the current number of credits for any user (via Credit Map) and how it should be updated (via Rate Map). Figure~\ref{fig:controller} shows these modules along with the data structures they manage.

The slice allocator intercepts resource requests from users, periodically executes the \name resource allocation algorithm (Algorithm~\ref{algo:karma-algorithm}) to compute allocations based on the user demands, and updates slices in the \mp accordingly. It interacts with the credit tracker to query and update user credits. A na\"ive implementation of Algorithm~\ref{algo:karma-algorithm} runs in $O(n\cdot f \cdot \log{n})$ time, where $n$ is the number of users, and $f$ is the fair share\footnote{The loop in Line~\ref{line:while-loop} of Algorithm~\ref{algo:karma-algorithm} takes $O(n\cdot f)$ iterations and each iteration would take $O(log n)$ time to find the donor/borrower with the minimum/maximum credits (if we were to maintain min/max heaps for the donor and borrower sets).}. Instead of computing allocations one slice at a time, we use an optimized implementation that carefully computes them in a batched fashion. This enables the slice allocator to support resource allocation at fine-grained timescales.

\paragraphb{Consistent hand-off of resources}
Since users are allowed to directly access slices from resource servers, we need to ensure consistent hand-off of slices from one user to another when slices are reallocated. For example, say user $U_1$ has a slice during a given quantum, and in the next quantum, this slice is allocated to user $U_2$. We need to ensure that (1) $U_1$'s data is flushed to persistent storage before $U_2$ overwrites it (2) $U_1$ should not be able to read/write to the slice after $U_2$ has accessed it (for example, there could be in-flight read/write requests to the slice which were initiated before $U_1$ gets to know it's allocation changed).

\name ensures the above by maintaining a monotonically increasing sequence number and current userID for each slice, at both the controller (within the \mp) and the resource servers (as slice metadata). On slice allocation, its userID is updated and its sequence number is incremented at the controller, and the sequence number is returned to the user. Subsequent user reads and writes to the slice specify this userID and sequence number. A slice read succeeds only if the accompanying sequence number is the same as the current slice sequence number, while a slice write succeeds only if the accompanying sequence number is the same or greater than the current sequence number. If a write necessitates an overwrite of the current slice content and metadata, the old slice content is transparently flushed persistent storage (\eg, S3) before the overwrite. In our example above, $U_2$'s first access to the slice after re-allocation will trigger a flush of $U_1$'s data to S3 and update the slice sequence number. Following this $U_1$'s accesses to this slice will fail since the current sequence number of the slices is higher. $U_1$ can then read/write this data from persistent storage.
Implementing consistent resource hand-off in Jiffy required minor changes to the controller (to track sequence numbers per slice), memory servers (to perform sequence number checking), and the client library (to tag requests with sequence numbers).

\begin{figure*}
  \centering
  \includegraphics[width = 1.0\textwidth]{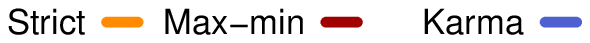}
    \subfigure[CDF of throughput across users] {
      \includegraphics[width = 0.25\textwidth]{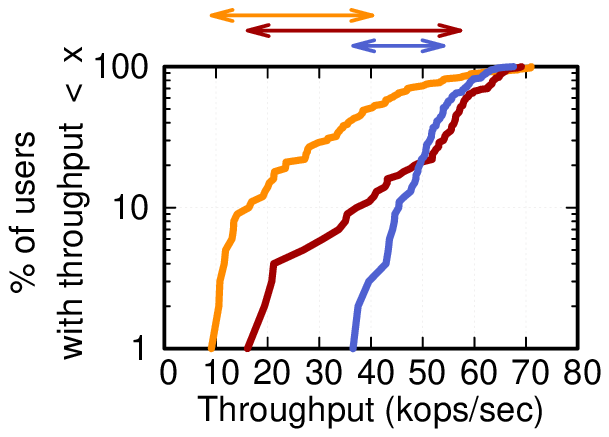}
      \label{fig:varyg-perf-cdf}
    }%
    \subfigure[CCDF of average latency across users] {
      \includegraphics[width = 0.25\textwidth]{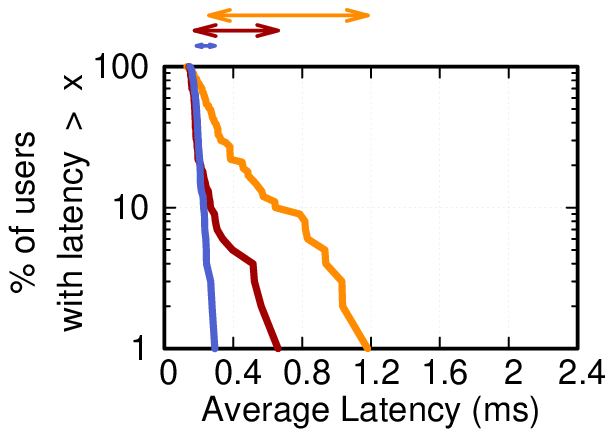}
      \label{fig:varyg-avglat-cdf}
    }%
    \subfigure[CCDF of P99.9 latency across users] {
      \includegraphics[width = 0.25\textwidth]{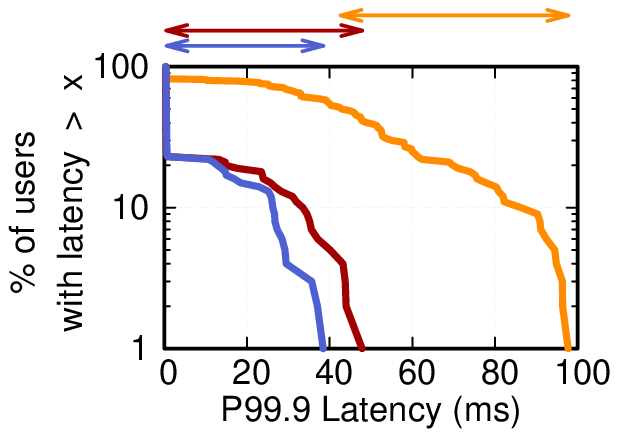}
      \label{fig:varyg-p999-cdf}
    }
    \subfigure[Disparity in throughput across users (median/min)] {
      \includegraphics[width=0.225\textwidth]{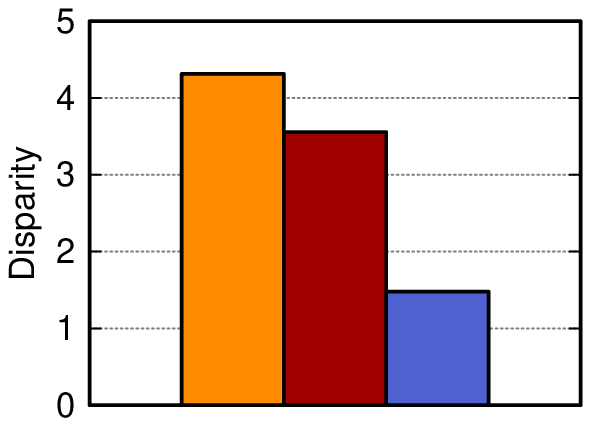}
      \label{fig:varyg-disparity}
    }
    \subfigure[Fairness in overall allocations (min/max allocation)] {
     \includegraphics[width = 0.225\textwidth]{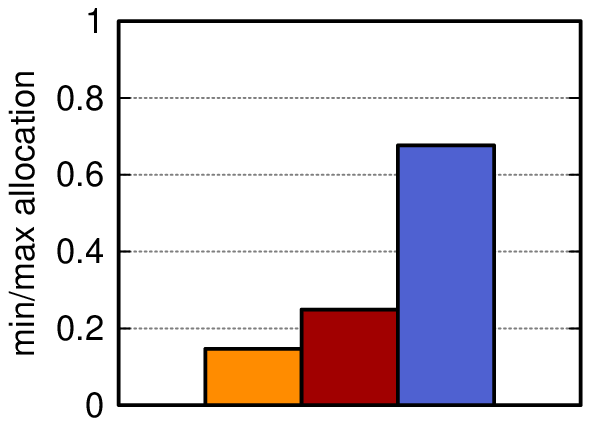}
     \label{fig:varyg-fairness}
    }
    \subfigure[System-wide average throughput (million ops/sec)] {
     \includegraphics[width = 0.225\textwidth]{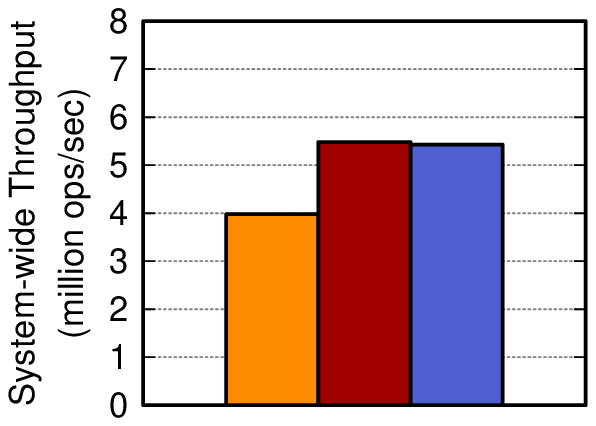}
     \label{fig:varyg-perf}
    }
  \vspace{-0.1in}
  \caption{\textbf{Understanding \name benefits.}
  (a) \name enables a much tighter throughput distribution across users (colored arrows show the absolute gap between median and minimum throughput across users). (b, c) It also enables a tighter distribution of average and tail latencies across users (again, colored arrows show the absolute gap between median and maximum latency across users). (d) \name achieves much lower throughput disparity---ratio of median to minimum values of throughput across users---than classic max-min fairness. (e) It also significantly reduces the gap between the users with minimum and maximum overall allocations, (f) while achieving similar system-wide performance as max-min fairness.}
  \label{fig:eval-fairness}
  \vspace{-0.05in}
\end{figure*}

  \vspace{-0.1in}
\section{Evaluation}
\label{sec:evaluation}
  \vspace{-0.05in}

We have already established \name properties theoretically in \S\ref{ssec:altruism}. In this section, we evaluate how \name's properties translate to application-layer benefits over an Amazon EC2 testbed with real-world workloads. Our evaluation demonstrates that:
\begin{denseitemize}
  \item \name reduces the performance disparity between different users by $\sim2.4\times$ relative to classic max-min fairness, without compromising on system-wide utilization or average performance (\S\ref{ssec:eval-dynamic});
  \item \name incentivizes users to share resources, quantifying \name's online strategy-proofness property (\S\ref{ssec:eval-unpredictable});
\end{denseitemize}

\noindent
We primarily focus on the shared cache use case from \S\ref{sec:overview} for the following reason. While datasets for the shared data analytics clusters use case are publicly available (\eg, Google and Snowflake datasets), they do not provide user queries that may impact our final conclusions. For the shared cache use case, we do have all the information we need: these datasets provide information on the working set size of each user over time, which can be fed into an end-to-end multi-tenant in-memory cache system running on Amazon EC2. We, thus, focus on this use case. 

\paragraphb{Experimental setup}
Our experimental setup consists of a distributed elastic in-memory cache shared across multiple users backed by a remote persistent storage system. For the cache, we use Jiffy~\cite{jiffy}, augmented with our implementation of \name (\S\ref{sec:arch}) and other evaluated schemes. If the evaluated scheme does not allocate sufficient slices to a user on Jiffy to fit its entire working set, the remaining data is accessed from remote persistent storage. When slices are reallocated between users across quanta, the corresponding data is moved between Jiffy and persistent storage through the consistent hand-off mechanism described in \S\ref{sec:arch}. We deployed our setup on Amazon EC2 using c5n.9xlarge instances ($36$ vCPUs, $96$GB DRAM, $50$Gbps network bandwidth). We host the Jiffy controller and resource servers across $7$ instances and use $25$ instances for the users/clients that issue queries to Jiffy. We use Amazon S3 as the persistent storage system.

\paragraphb{Workload}
We use the publicly available Snowflake dataset~\cite{snowset} that provides dynamic user demands in terms of memory usage for each customer from Snowflake's production cluster. We use these demands as the dynamic working set size for individual users. For each user, we issue data access queries using the standard YCSB-A workload~\cite{ycsb} (50\% read, 50\% write) with uniform random access distribution, with queries during each quantum being sampled (according to the YCSB parameters) within the instantaneous working set size of that user. If a query references data that is currently cached in Jiffy, then it is serviced directly from the corresponding resource server; otherwise, it is serviced from the persistent storage. 

\paragraphb{Default parameters}
Unless specified otherwise, we randomly choose $100$ users (out of $\sim2000$ users) over a randomly-chosen $15$ minute time window (out of a 14-day period) in the Snowflake workload. To test for extreme scenarios, we set the length of each quantum to be one second (that is, a total of $900$ quanta). The fair share of each user is $10$ slices, and the total memory capacity of the system is set to the number of users times the fair share ($1000$ slices). Each slice is $128$MB in size, while each query corresponds to a read or write to a $1$KB chunk of data (the default size in the YCSB workload).

\paragraphb{Compared schemes} 
We compare Karma to strict partitioning and max-min fairness, since they correspond to the two most popular fair allocation schemes, and represent extremes in resource allocation and performance. When evaluating \name, we set the number of initial credits to a large value\footnote{As discussed in \S\ref{ssec:algo-discussion}, the precise value is unimportant. Here, we set it to $900,000$, so that even if a user was allocated the full system capacity for the entire duration ($1000 \times 900$) it would not run out of credits.}. The fraction of fair share that is guaranteed ($\alpha$) is $0.5$ by default.

\begin{figure*}
  \centering
  \subfigure[Utilization] {
     \includegraphics[width = 0.225\textwidth]{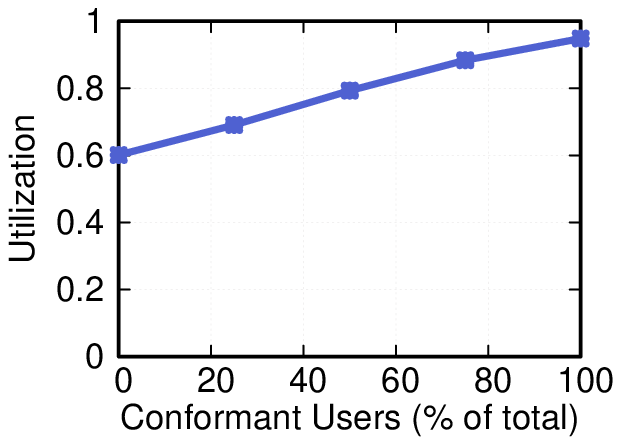}
     \label{fig:varyalt-util}
   }
   \subfigure[Performance] {
     \includegraphics[width = 0.225\textwidth]{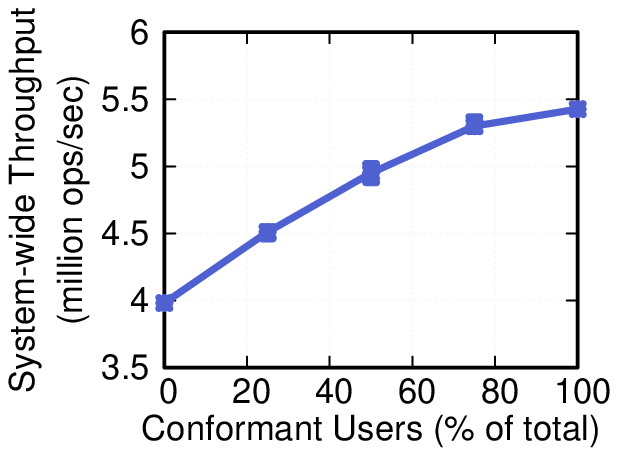}
     \label{fig:varyalt-perf}
   }
   \subfigure[Welfare improvement] {
     \includegraphics[width = 0.225\textwidth]{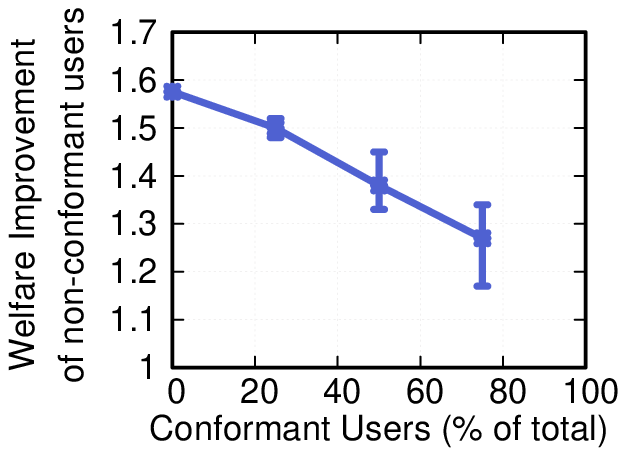}
     \label{fig:varyalt-incentive}
   }%
   \vspace{-1em}
  \caption{\textbf{\name incentivizes resource sharing.} All metrics are computed as averages (with error bars) for three random selections of users being non-conformant. See \S\ref{ssec:eval-unpredictable} for details.}
  \label{fig:eval-main}
\end{figure*}

\paragraphb{Metrics} We evaluate system-wide resource utilization, along with both per-user and system-wide performance---key metrics for any resource allocation mechanism. For performance, we measure both throughput and latency (average and 99.9th percentile tail). We define performance \emph{disparity} for an allocation scheme as the ratio of median to minimum performance (that is, throughput or latency) observed across various users. For any given user, we define \textit{welfare} over time  $t$ as $\frac{\sum_{t} \text{allocations}}{\sum_{t} \text{demands}}$, that is, the fraction of its total demands satisfied by the allocation scheme.  We define \textit{fairness} as $\frac{min_{users} \text{welfare}}{max_{users} \text{welfare}}$ (higher is better, $1$ is optimal), as a measure of welfare disparity between users.

\vspace{-0.1in}
\subsection{Understanding \name Benefits}
\label{ssec:eval-dynamic}
\vspace{-0.05in}
We now evaluate \name's benefits in terms of reducing disparity across users' application-level performance as well as resource allocation.

\paragraphb{\name reduces performance disparity between users} Figure~\ref{fig:varyg-perf-cdf} shows the throughput distribution across users for our compared schemes; the y-axis is presented in log-scale to focus on the  users at the tail of the distribution, which observe the most performance disparity. Since \name strives to balance fairness over time, it significantly narrows the throughput distribution across users compared to the two baselines: the ratio between the maximum and minimum throughput across all users is $7.8\times$ with strict partitioning and $4.3\times$ with max-min fairness, but only 1.8$\times$ for \name. As Figure~\ref{fig:varyg-disparity} shows, \name lowers the throughput disparity across users by $2.4\times$ compared to max-min fairness. \name also reduces average latency disparity (Figure~\ref{fig:varyg-avglat-cdf}) by $2.4\times$ and 99.9th percentile latency disparity (Figure~\ref{fig:varyg-p999-cdf}) by $1.2\times$ compared to max-min fairness by enabling a tighter distribution for both latencies.

Equitability in performance across users for a scheme is closely tied to how fairly resources are allocated across users. Specifically, because of the large gap between elastic memory (Jiffy) and S3 latencies ($50$--$100\times$), accesses to slices in S3 result in significantly lower throughput than accesses to slices in elastic memory. As a result, users' average throughput ends up being roughly proportional to their total allocation of slices in elastic memory over time. Similarly, since a larger total allocation results in a smaller fraction of requests going to S3, average and tail latencies also reduce.

\paragraphb{\name reduces disparity in allocations}
We now quantify disparities in overall allocations obtained by users across our compared schemes via our fairness metric in Figure~\ref{fig:varyg-fairness}. Due to dynamic demands, strict partitioning exhibits very poor fairness, since users with very bursty demands end up getting much lower total allocations than users who have steady demands\footnote{Note that only \textit{useful} allocations are considered---strict partitioning guarantees a fixed allocation at all times, but resources may remain unused when demand is low.}. While, max-min fairness observes better fairness compared to strict partitioning, the best-off user still receives $4\times$ higher allocation than the worst-off user, resulting in poor absolute fairness. \name achieves significantly better fairness with the best-off user receiving only $1.5\times$ higher allocation than the worst-off user. It is able to achieve this by prioritizing the allocation of resources beyond the fair share to users with more credits (\S\ref{ssec:adr}).

\paragraphb{\name achieves Pareto efficiency and high system-wide performance}
\name achieves the same overall resource utilization as max-min fairness ($\sim 95\%$). This is because \name is Pareto efficient (\S\ref{ssec:formal}) similar to max-min fairness and thus achieves near-optimal utilization. We find that the optimal utilization is $<100\%$ since some quanta observe total user demands less than system capacity.

Max-min fairness observes $1.4\times$ higher system-wide throughput (that is, throughput aggregated across all users) than strict partitioning (Figure~\ref{fig:varyg-perf}) since it permits allocations beyond the fair share, allowing more requests to be served on faster elastic memory. \name observes system-wide performance similar to max-min fairness for similar reasons; the slight variations are attributed to variance in S3 latencies.

\begin{figure*}
  \centering
  \subfigure[Utilization] {
    \includegraphics[width = 0.25\textwidth]{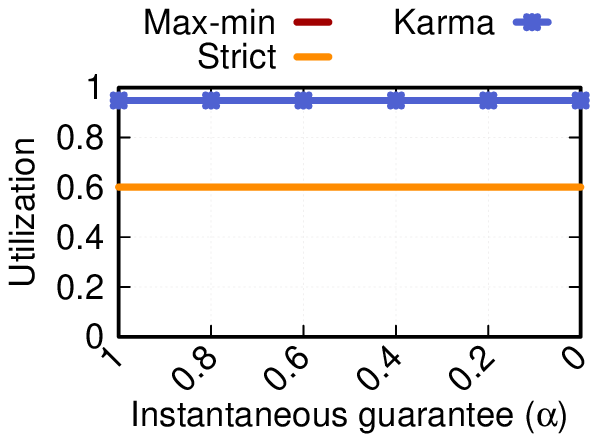}
    \label{fig:varyg-sensitivity-util}
  }
  \subfigure[Performance]{
    \includegraphics[width = 0.25\textwidth]{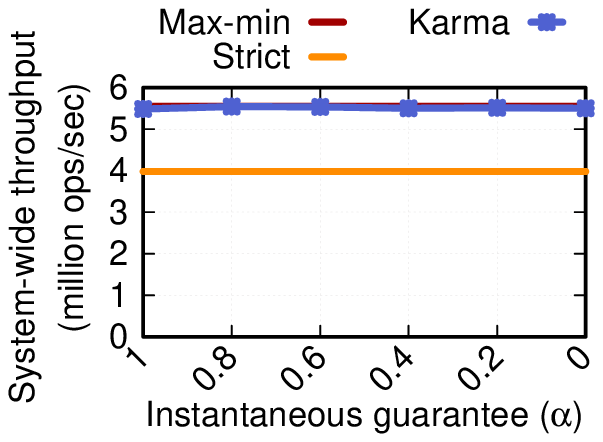}
    \label{fig:varyg-sensitivity-perf}
  }
  \subfigure[Fairness]{
    \includegraphics[width = 0.25\textwidth]{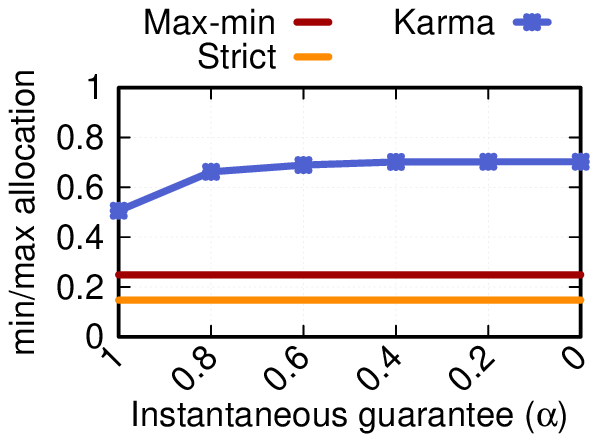}
    \label{fig:varyg-sensitivity-fairness}
  }
  \vspace{-1em}
  \caption{\textbf{Sensitivity analysis with varying instantaneous guarantee ($\alpha$)} (a, b) \name matches the resource utilization and system-wide performance of max-min fairness independent of $\alpha$ (c) Smaller values of $\alpha$ result in improved long-term fairness.}
  \label{fig:varyg-sensitivity}
\end{figure*}

\subsection{\name Incentives}
\label{ssec:eval-unpredictable}
\vspace{-0.05in}
We now empirically demonstrate that \name incentivizes users to donate resources instead of hoarding them, to improve their own as well as overall system welfare. To this end, we vary the fraction of users using \name that are \textit{conformant} or \textit{non-conformant}. A conformant user is truthful about its demands and donates its resources when its demand is less than its fair share. A non-conformant user, on the other hand, always asks for the maximum of its demand or its fair share (that is, it over-reports its demand during some quanta).

\paragraphb{Resource utilization and system-wide performance improve with more conformant users} Figure~\ref{fig:varyalt-util} and Figure~\ref{fig:varyalt-perf} show that \name's system-wide utilization and performance improve as the fraction of conformant users increases. This is because as more users donate resources when they do not need them, other users can use these resources, improving overall utilization and performance. When none of the users are conformant, since no one ever donates any resources, \name essentially reduces to strict partitioning, hence achieving low overall utilization and performance. When all users are conformant, \name achieves optimal utilization and performance, similar to classic max-min fairness.

\paragraphb{Becoming conformant improves user welfare} Figure~\ref{fig:varyalt-incentive} shows the average welfare gain non-conformant users would achieve if they were to become conformant. When non-conformant users become conformant, it leads to significant ($1.17$--$1.6\times$) welfare gains for them, empirically validating \name's property that users have nothing to gain by over-reporting their demand (\S\ref{ssec:formal}). Note that the gain varies with the number of conformant users in the system---the gains from non-conformant users becoming conformant are higher when the percentage of conformant users is low. As expected, the gains show diminishing returns as more users in the system become conformant as overall utilization is already high. 

\subsection{\name Sensitivity Analysis}
\label{ssec:eval-sensitivity}

We now show sensitivity analysis with the only parameter in the \name algorithm--the instantaneous guarantee ($\alpha$). Figure~\ref{fig:varyg-sensitivity} shows the resource utilization, system-wide performance, and fairness with $\alpha$ varying between $0$ and $1$. \name continues to match the resource utilization and system-wide performance of max-min fairness independent of $\alpha$ (Figure~\ref{fig:varyg-sensitivity-util} and Figure~\ref{fig:varyg-sensitivity-perf}). Varying $\alpha$ has an impact on the long-term fairness achieved by \name (Figure~\ref{fig:varyg-sensitivity-fairness}), with smaller values of $\alpha$ resulting in improved fairness, thus validating our discussion in \S\ref{ssec:algo-discussion}. Even for $\alpha = 1$, \name is able to achieve significantly better fairness compared to max-min fairness. This is because, while it allocates resources up to the fair share identically to max-min fairness, it prioritizes allocation beyond the fair share based on credits.

  \vspace{-0.1in}
\section{Related Work}
\label{sec:related}
  \vspace{-0.05in}

There is a large and active body of work on resource allocation and scheduling, exploring various models and settings; it would be a futile attempt to compare \name with each individual work. We do not know of any other resource allocation mechanism that guarantees Pareto efficiency, strategy-proofness, and fairness similar to \name for the case of dynamic user demands; nevertheless, we discuss below the most closely related works.

\paragraphb{Max-min fairness variants in cloud resource allocation and cluster scheduling}
Many works study variants of max-min fairness for cloud resource allocation and cluster scheduling~\cite{dpf,carbyne,drf,pisces,faircloud,fairride,graphene,tetris,hug,elasticswitch,seawall,rrf,hadrian,pulsar,proteus}, including recent work on ML job scheduling~\cite{gandiva-fair,tiresias,optimus,themis,gavel}. We make three important notes here.
First, while dominant resource fairness (DRF)~\cite{drf} has generalized max-min fairness to multiple resources, it makes the same assumptions as max-min fairness: user demands being static over time; our goals are different: we have identified and resolved the problems with max-min fairness for the case of a single resource but over dynamic user demands. It is an interesting open problem to generalize \name for the case of multiple resources.

Second, cluster scheduling has been studied under several metrics beyond fair resource allocation (\eg, job completion time, data locality, priorities, etc.). Themis~\cite{themis} considers long-term fairness but defines a new ML workload-specific notion of fairness, and is therefore not directly comparable to \name. Our goals are most aligned with those works that study fair allocation under strategic users while guaranteeing Pareto efficiency. To that end, the closest to \name is \textsc{Carbyne}~\cite{carbyne}. However, \textsc{Carbyne} not only assumes non-strategic users but also, for the single-resource case (the focus of this paper), \textsc{Carbyne} converges to max-min fairness. As discussed earlier, generalizing \name to multiple resources remains an open problem; a solution for that problem must be compared against \textsc{Carbyne}.

Finally, fairness in application-perceived performance is only indirectly related to fairness in resource allocation: other factors like software systems (\eg, hypervisors and storage systems) and resource preemption granularity can impact performance. Similar to other mechanisms~\cite{snowset,atikoglu2012workload, carbyne, pisces,graphene,tetris, drf, stoica1996proportional, dpf, hadrian, proteus, alibabatrace1, googletrace, alibabatrace2, fairride, delaysc, quincy}, \name's properties are independent of these system-level factors; while our evaluation shows that \name properties translate to application-level benefits, absolute numbers depend on the underlying system implementation.

\paragraphb{Allocation of time-shared resources}
Generalized Processor Sharing (GPS)~\cite{gps} is an idealized algorithm for sharing a network link which assumes that traffic is infinitesimally divisible (fluid model). For equal-sized packets and equal flow weights, GPS reduces to Uniform Processor Sharing~\cite[Section 2]{gps}, which is equivalent to max-min fairness. GPS guarantees fairness over arbitrary time intervals only under the assumption that flows are {\em continuously backlogged} \cite[Section 2]{gps}. This assumption implies that flows always have demand greater than their fair share, making it trivial to guarantee a max-min fair share of the network bandwidth over arbitrary time intervals. Classical fair-queueing algorithms~\cite{wfq,drr,wf2q,sfq,wf2qplus-journal} in computer networks approximate GPS with the constraint of packet-by-packet scheduling. Under this constraint, varying-sized packets and different flow weights make it hard to realize fairness efficiently; thus, the technical question that these algorithms solve is to achieve fairness approximately equal to GPS with minimal complexity. \name focuses on a different problem---we show that GPS guarantees (equivalent to max-min fairness) are not sufficient when demands are dynamic and present new mechanisms to achieve fairness while maintaining other properties for such dynamic demands.

Stride~\cite{stride-scheduling} scheduling essentially approximates GPS in the context of CPU scheduling \cite[Section 7]{stride-scheduling}, and thus the above discussion applies to it as well. DRF-Q~\cite{drf-q} generalizes DRF to support both space and time-shared resources, but is explicitly designed to be memoryless similar to max-min fairness, and therefore suffers from similar issues for long-term fairness. Least Attained Service (LAS)~\cite{las-survey,las-bianca,las-kleinrock} is a classical job scheduling algorithm that has been applied to packet scheduling~\cite{las-bianca}, GPU cluster scheduling~\cite{tiresias}, and memory controller scheduling~\cite{atlas}. 
For $\alpha=0$, \name behaves similarly to LAS, and for $\alpha > 0$, \name generalizes LAS with instantaneous guarantees. Moreover, our results from \S\ref{ssec:formal} establish strategy-proofness properties of LAS for dynamic user demands, which may be of independent interest.

\paragraphb{Theory works}
Several recent papers in the theory community study the problem of resource allocation for dynamic user demands. Freeman et al.~\cite{freeman-zcl} and Hossain et al.~\cite{hossain} consider dynamic demands under a different setting, where users can benefit when they are allocated resources above their demand; under this setting, they focus on instantaneous fairness (which is non-trivial since users can be allocated resources beyond their demand). Karma instead focuses on long-term fairness under the traditional model, where users do not benefit from resources beyond their demands. Sadok et al.~\cite{sadok-cc21} present minor improvements over max-min fairness for dynamic demands. Their mechanism allocates resources in a strategy-proof manner according to max-min fairness while marginally penalizing users with larger past allocations using a parameter $\delta\in [0,1)$. For both $\delta=0$ and $\delta\to 1$, the penalty goes to $0$ for every past allocation, and the mechanism becomes identical to max-min fairness; for other values of $\delta$, the penalty is at most a $\delta(1-\delta) \le 1/4$ fraction of past allocation surplus, and it reduces exponentially with time (users who were allocated large amounts of resources further in the past receive an even smaller penalty). Thus, for all values of $\delta$, and in particular, for $\delta=0$ and $\delta \to 1$, their mechanism suffers from the same problems as max-min fairness. Aleksandrov et al.~\cite{aleksandrov2019strategy} and Zeng et al.~\cite{engP20} consider dynamic demands, but in a significantly different setting than ours where resources arrive over time.

\paragraphb{Pricing- and credit-based resource allocation} Another stream of work related to \name is pricing-based and bidding-based mechanisms for resource allocation, \eg, spot instance marketplace and virtual machine auctions~\cite{spot-market,zheng2015bid,zheng2016viability,wolski2017probabilistic,ginseng1,ginseng2}. While interesting, this line of work does not focus on fair resource allocation and is not applicable to use cases that \name targets.
XChange~\cite{xchange} proposes a market-based approach to fair resource allocation in multi-core architectures but focuses on instantaneous fairness rather than long-term fairness, unlike \name. 
It assigns a ``budget'' of virtual currency to each user which can be used to bid for resources. This budget is however reset during every time quantum, and therefore information about past allocations is not carried over.

Credits are used in many other game theoretic contexts~\cite{nisan2001algorithmic,feigenbaum2004distributed,roughgarden2010algorithmic}, \eg, in peer-to-peer and cooperative caching settings to incentivize good behavior among participants with static demands~\cite{piatek2007incentives,yadgar2013cooperative,cox2003samsara}. However, we are not aware of any credit-based mechanisms that deal with resource allocation in the context of dynamic user demands.

\vspace{-0.1in}
\section{Conclusion}
\vspace{-0.05in}
This paper builds upon the observation that the classical max-min fairness algorithm for resource allocation loses one or more of its desirable properties---Pareto efficiency, strategy-proofness, and/or fairness---for the realistic case of dynamic user demands. We present \name, a new resource allocation mechanism for dynamic user demands, and theoretically establish \name guarantees related to Pareto efficiency, strategy-proofness, and fairness for dynamic user demands. Experimental evaluation of a realization of \name in a multi-tenant elastic memory system demonstrates that \name's theoretical properties translate well into practice: it reduces application-level performance disparity by as much as $2.4\times$ when compared to max-min fairness while maintaining high resource utilization and system-wide performance.

\name opens several exciting avenues for future research. These include (but are not limited to) extending \name theoretical analysis for $\alpha > 0$, generalizing \name to allocate multiple resource types (similar to DRF), extending \name to handle all-or-nothing or gang-scheduling constraints which are prevalent in the context of GPU resource allocation~\cite{gandiva-fair,themis}, and applying \name to other use cases such as inter-datacenter network bandwidth allocation and resource allocation for burstable VMs in the cloud.

\vspace{-0.1in}
\section*{Acknowledgements}
\vspace{-0.05in}
We thank our shepherd, Sebastian Angel, and the OSDI reviewers for their insightful feedback. This research was supported in part by NSF CNS-$1704742$, CNS-$2047220$, CNS-$2047283$, CNS-$2104292$, CNS-$2143868$, AFOSR grants FA$9550$-$19$-$1$-$0183$, FA$9550$-$23$-$1$-$0068$, a NetApp Faculty Fellowship, an NDSEG fellowship, a Sloan fellowship, and gifts from Samsung, VMware, and Enfabrica.
\end{sloppypar}

\bibliographystyle{plain}
\bibliography{bib/paper}
\clearpage{
\appendix
\section{Proofs}
\label{appendix}
\subsection{Proof for worst-case max-min fairness}
\label{max-min-bad}
\begin{theorem}
  There is an instance with $n$ users where the ratio of the maximum over the minimum resource allocation in classical max-min fairness is $n+1$.
\end{theorem}

\begin{proof}
  The amount of available resources every quantum is $n$, the fair share of every user is $f=1$, and there are $n+1$ quanta in total. The first $n-1$ users constantly have demand $1$ and the final user, A, has $0$ demand the first $n$ quanta, but demand $n$ on the final quantum. Classic max-min fairness will always allocate $1$ resource to every one of the $n-1$ users, leading to a total allocation of $n+1$ in the final quantum, while on the same quantum user A will have received a total allocation of $1$. In this scenario, the ratio of minimum over maximum allocation is $n+1$.
\end{proof}

\subsection{Strategy-proofness analysis preliminaries}
\label{app:proofs}

We denote $x^+ = \max(x,0)$. 

The set of users is denoted with $[n]$. 
We denote with $r_i^t$ the allocation of user $i$ in quantum $t$. We also denote with $R_i^t$ the cumulative allocation of user $i$ up to round $t$, i.e. $R_i^t = \sum_{\tau=1}^t r_i^\tau$. By definition, $R_i^0=0$. Since we focus on the $\alpha = 0$ case, we note that the number of credits user $i$ earns in quantum $t$ is $f - r_i^t$. Using the notation of this section, this means that \name prioritizing users with the most credits is equivalent to prioritizing users with the least cumulative allocation.

Every quantum $t$, each user $i$'s real demand is $d_i^t$. The useful allocation of user $i$ on quantum $t$ is $u_i^t = \min(r_i^t, d_i^t)$. The total useful allocation of user $i$ after quantum $t$ equals the sum of useful allocations up to that round, i.e. $U_i^t = \sum_{\tau=1}^t u_i^\tau = \sum_{\tau=1}^t \min(r_i^\tau, d_i^\tau)$.

W.l.o.g. we are usually  going to study the possible deviations of user $1$, i.e. how much user $1$ can increase its allocation by lying about its demand. We use the symbols $\hat d_i^t$, $\hat r_i^t$, $\hat R_i^t$, $\hat u_i^t$, $\hat U_i^t$ to denote the claimed demand and resulting outcome of some deviation of user $1$.

We first prove a couple of auxiliary lemmas.

\begin{lemma}\label{lem:single:more_less}
    Fix a quantum $t$ and let $i, j$ be two different users. If the following conditions hold
    \begin{itemize}
        \item $r_i^t < \hat r_i^t$ and $r_i^t < d_i^t$, i.e. user $i$ gets more resources on quantum $t$ when user $1$ deviates and user $i$ could have gotten more resources when user $1$ does not deviate.
        
        \item $r_j^t > \hat r_j^t$ and $\hat r_j^t < \hat d_j^t$, i.e. user $j$ gets less resources on quantum $t$ when user $1$ deviates and user $j$ could have gotten more resources when user $1$ deviates.
    \end{itemize}
    then $R_i^t \geq R_j^t$ and $\hat R_i^t \le\hat R_j^t$, implying
    \begin{equation}\label{eq:single:0}
        \hat R_i^t - R_i^t \le \hat R_j^t - R_j^t
    \end{equation}
\end{lemma}

It should be noted that the conditions for $i$ (similarly for $j$) can be simplified if $i$ has the same demand in both outcomes (which is trivially true if $i\neq 1$): if $r_i^t < \hat r_i^t$ the other inequality is implied as $\hat d_i^t = d_i^t$ and $\hat r_i^t \le \hat d_i^t$.

\begin{proof}
    Because of the conditions, we notice that $r_i^t < d_i^t$ and $r_j^t > 0$ (the last inequality is true because of $\hat r_j^t < \hat d_j^t$ which guarantees that $\hat r_j^t \ge 0$ and $r_j^t > \hat r_j^t$). This implies that it would have been feasible to increase $r_i^t$ by decreasing $r_j^t$.
    This implies that $R_i^t \ge R_j^t$; otherwise \name would have prioritized user $j$ and given it some of the resources user $i$ got. 
    With the analogous inverse argument (we can increase $\hat r_j^t$ by decreasing $\hat r_i^t$) we can prove that $\hat R_i^t \le \hat R_j^t$. This completes the proof.
\end{proof}

The main technical tool in our work is the following lemma bounding the total amount all the users have ``won'' because of user $1$ deviating, i.e. $\sum_k(\hat R_k^t - R_k^t)^+$. So rather than bounding the deviating user $1$'s gain directly, it is better to consider the overall increase in all users combined.
More specifically, the lemma upper bounds the increase of that amount after any quantum, given that user $1$ does not over-report her demand.  The bound on the total over-allocation $\sum_k(\hat R_k^t - R_k^t)^+$ then follows by summing over the time periods. The bound on the increase of $\sum_k(\hat R_k^t - R_k^t)^+$ after any quantum is different according to three different cases: 
\begin{itemize}
    \item If all users' demands are satisfied, then the increase is at most $0$.
    \item If user $1$ is truthful the increase is again at most $0$ so in these steps over-allocation can move between users but cannot increase. This is the reason working with the total over-allocation is so helpful.
    \item If user $1$ under-reports the increase is bounded by the amount of resources she receives when she is truthful.
\end{itemize}

\begin{lemma}\label{lem:single:aux_bound}
    Fix any $t\ge 1$. Let $\{R_i^{t-1}\}_{i\in [n]}$ and $\{\hat R_i^{t-1}\}_{i\in [n]}$ be the cumulative allocations up to quantum $t-1$. Assume that $\{d_i^t\}_{i\in [n]}$ are some users' demands and that $\{\hat d_i^t\}_{i\in [n]}$ are the same demands except user $1$'s, who deviates but does not over-report, i.e. $\hat d_1^t \le d_1^t$. Then it holds that
    \begin{equation}\label{eq:single:aux}
    \begin{split}
        \sum_{k\in [n]}\left(\hat R_k^t - R_k^t\right)^+
        &-
        \sum_{k\in [n]}\left(\hat R_k^{t-1} - R_k^{t-1}\right)^+ \\
        &\le
        r_1^t \One{\hat d_1^t < d_1^t}
    \end{split}
    \end{equation}
\end{lemma}

When all demands are satisfied, user $1$ clearly cannot change other users' allocations by under-reporting. We will use \Cref{lem:single:more_less} to show that if user $1$ is truthful on quantum $t$, then the l.h.s. of \eqref{eq:single:aux} is at most $0$; as the mechanism allocates resources such that the large $\hat R_i^t - R_i^t$ are decreased and the small $\hat R_i^t - R_i^t$ are increased.  Finally, if user $1$ under-reports its demand then the (at most) $r_1^t$ resources user $1$ does not get might increase the total over-allocation by the same amount. 

\begin{proof}
    Define $P^t=\{i\in [n]: \hat R_i^t \ge R_i^t\}$ for all $t$. Suppose by contradiction:
    \begin{equation*}
        \sum_{k\in P^t}\left(\hat R_k^t - R_k^t\right)
        -
        \sum_{k\in P^{t-1}}\left(\hat R_k^{t-1} - R_k^{t-1}\right)
        >
        r_1^t \One{\hat d_1^t < d_1^t}
    \end{equation*}
    
    Because $\sum_{k\in P^{t}}(\hat R_k^{t-1} - R_k^{t-1})\le\sum_{k\in P^{t-1}}(\hat R_k^{t-1} - R_k^{t-1})$, the above inequality implies
    \begin{equation}\label{eq:single:11}
        \sum_{k\in P^t}\left(\hat r_k^t - r_k^t\right) > r_1^t \One{\hat d_1^t < d_1^t}
    \end{equation}

    Because user $1$ does not over-report its demand, it holds that $\sum_k r_i^t \ge \sum_k \hat r_i^t$, i.e. the total resources allocated to the users does not increase when user $1$ deviates. Combining this fact with \eqref{eq:single:11} we get that
    \begin{equation}\label{eq:single:12}
        \sum_{k\notin P^t}\left(r_k^t - \hat r_k^t\right) > r_1^t \One{\hat d_1^t < d_1^t}
    \end{equation}

    We notice that because of \eqref{eq:single:11}, there exists a user $i\in P^t$ for whom $\hat r_i^t > r_i^t$; because of \eqref{eq:single:12}, there exists a user $j\notin P^t$ for whom $r_j^t > \hat r_j^t$. Additionally for that $j$ we can assume that $\hat d_j^t = d_j^t$ because:
    \begin{itemize}
        \item If user $1$ does not deviate then for all $k$, $\hat d_k^t = d_k^t$.
        
        \item If $\hat d_1^t < d_1^t$, then \eqref{eq:single:12} implies $\sum_{k\notin P^t,\,k\neq 1}\left(r_k^t - \hat r_k^t\right) > 0$, i.e. $j\neq 1$ and we assumed that only user $1$ deviates.
    \end{itemize}
    
    Thus we have $\hat d_i^t \le d_i^t$ (since no user over-reports), $\hat d_j^t = d_j^t$, $\hat r_i^t > r_i^t$, and $\hat r_j^t < r_j^t$. Now \Cref{lem:single:more_less} proves that $\hat R_i^t - R_i^t \le \hat R_j^t - R_j^t$. This leads to a contradiction, because $i\in P^t$ and $j\notin P^t$, i.e. $\hat R_i^t - R_i^t \ge 0 > \hat R_j^t - R_j^t$.
\end{proof}

\subsection{Proof of Lemma 1 -- Over-reporting}
\label{app:lem1p}
Now we prove Lemma 1 from the main paper, that users have no incentive to over-report their demand.

\begin{proof}[Proof of Lemma 1]
    Fix a quantum $T$ and let $\{\hat d_i^t\}_{i,t}$ be any demands (that possibly involve user $1$ both over and under-reporting). We are going to show that if user $1$ changes every over-report to a truthful one, then its utility on quantum $T$ is not going to decrease.  Let $T_0\le T$ be the last quantum where user $1$ over-reported. 
    For all users $i$ and quanta $t\in[1,T]$, let $\bar d_i^t = \hat d_i^t$, except for $\bar d_1^{\,T_0}$ which is $1$'s actual demand (note that $\bar d_1^{\,T_0} < \hat d_1^{T_0}$). Let $\bar r$, $\bar R$, $\bar u$, $\bar U$ be the result of demands $\bar d$. We will show that $\bar U_1^T \ge \hat U_1^T$, i.e.  user $1$ does not prefer the demands $\{\hat d_i^t\}_{i,t}$ over the demands $\{\bar d_i^t\}_{i,t}$.  If we apply this inductively for every quantum before $T$ where user $1$ over-reports, we are going to get that over-reporting is not a desirable strategy.
    
    Up to quantum $T_0-1$ all users' demands in $\bar d$ and $\hat d$ are the same and thus so are the allocations: for all $i$, $\hat R_i^{T_0-1} = \bar R_i^{T_0-1}$ and $\hat U_i^{T_0-1} = \bar U_i^{T_0-1}$. Because $\hat d_1^{T_0} > \bar d_1^{\,T_0}$ and for $i\neq 1$, $\hat d_i^{T_0} = \bar d_i^{\,T_0}$, user $1$ may earn some additional resources on $T_0$, i.e. $\hat r_1^{T_0} - \bar r_i^{T_0} = \hat R_1^{T_0} - \bar R_i^{T_0} = x$, for some $x\ge 0$, while other users $i\neq 1$ get less resources: $\hat r_i^{T_0} - \bar r_i^{T_0} = \hat R_i^{T_0} - \bar R_i^{T_0} \le 0$. We first note that the $x$ additional resources that user $1$ gets are in excess of $1$'s true demand, meaning they do not contribute towards $1$'s useful allocation:
    \begin{equation}\label{eq:single:1}
        \hat U_1^{T_0} - \bar U_1^{T_0} = \hat R_1^{T_0} - \bar R_1^{T_0} - x = 0
    \end{equation}
    
    Additionally, because user $1$ does not over-report $\hat d$ or $\bar d$ in quanta $T_0+1$ to $T$ (by assumption $T_0$ is the last quantum before $T$ where user $1$ over-reports), it holds that for $t\in [T_0+1, T]$, user $1$'s useful allocation is the same as the resources it receives: $\bar u_1^t = \bar r_1^t$ and $\hat u_1^t = \hat r_1^t$.  This fact, combined with \eqref{eq:single:1} proves that
    \begin{equation}
        \forall t\in[T_0,T],\;
        \hat U_1^t - \bar U_1^t = \hat R_1^t - \bar R_1^t - x
    \end{equation}
    
    Thus, in order for this over-reporting to be a strictly better strategy, it most hold that $\hat R_1^T - \bar R_1^T > x$. We will complete the proof by proving that the opposite holds. Since there is no over-reporting in periods $t\in[T_0+1, T]$ we can use \Cref{lem:single:aux_bound}, where we use $\hat d_i^t\le \bar d_i^t$ in place of $\hat d_i^t\le d_i^t$ and summing \eqref{eq:single:aux} for all $t\in [T_0+1, T]$ and noticing that $\hat d_1^t = \bar d_1^t$ we get 
    \begin{equation*}
        \sum_k\left(\hat R_k^T - \bar R_k^T\right)^+
        -
        \sum_k\left(\hat R_k^{T_0} - \bar R_k^{T_0}\right)^+
        \le 0
    \end{equation*}
    
    The above, because $(\hat R_k^T - \bar R_k^T)^+\ge 0$, $\bar R_k^{T_0} - \bar R_k^{T_0} \le 0$ if $k\neq 1$, and $\bar R_1^{T_0} - \bar R_1^{T_0} = x \ge 0$, proves that $\hat R_1^T - \bar R_1^T \le x$. This completes the proof.
\end{proof}

\subsection{Proof of Online Strategy-proofness}
\label{ssec:osp}
We now prove that \name is online strategy-proof.
\begin{proof}[Proof of Online strategy proofness]
    Fix a quantum $t$ and assume that all users' demands are truthful up to quantum $t-1$. To prove that \name is online strategy-proof we need to prove that given these conditions, user $1$ cannot increase its demand on quantum $t$. Because over-reporting demand never leads to increased utility, user $1$ can increase its utility only by under-reporting. However, simply under-reporting (without altering past allocations) can only decrease user $1$'s allocation. This completes the theorem's proof.
\end{proof}

\subsection{Proof of Lemma 2}
\label{ssec:thm:more_than_triple}
Now we prove that users cannot misreport their demand to increase their resources by a factor larger than $1.5$ and that if they do so they might decrease their total useful allocation by a factor of $\frac{n+2}{2}$.

\begin{proof}[Proof of Lemma 2 -- upper bound on potential increase]
    Lemma 1 implies that it is no loss of generality to assume that user $1$ does not over-report its demand, since any benefit gained by over-reporting can be gained by changing every over-report to a truthful one. This means that instead of $\hat U_1^t \le 1.5 U_1^t$ we can show $\hat R_1^t \le 1.5 R_1^t$. Towards a contradiction, let $t$ be the first quantum when user $1$ gets more than $1.5$ more resources by some deviation of demands, i.e. 
   $\hat R_1^t > 1.5 R_1^t$ and $\hat R_1^{t-1} \le 1.5 R_1^{t-1}$. This implies that $\hat r_1^t > r_1^t$, which in turn entails that there exists a user $j$ for who $\hat r_j^t < r_j^t$, since the total resources allocated when user $1$ is under-reporting cannot be less than those when 1 is truthful. Because $\hat r_1^t > r_1^t$, $\hat r_j^t < r_j^t$, $\hat d_1^t \le d_1^t$, and $\hat d_j^t = d_j^t$, we can use \Cref{lem:single:more_less} and get $\hat R_1^t - R_1^t \le \hat R_j^t - R_j^t$. This inequality, $\hat R_1^t - R_1^t \geq 0$, and \Cref{lem:single:aux_bound} by summing \eqref{eq:single:aux} for every quantum up to $t$, implies
    \begin{align*}
        2\left(\hat R_1^t -  R_1^t\right)
        &\le
        \left(\hat R_1^t -  R_1^t\right) + \left(\hat R_j^t -  R_j^t\right)\\
        &\le
        \sum_k\left(\hat R_k^t -  R_k^t\right)^+
        \le
        \sum_{\tau=1}^t r_1^\tau = R_1^t
    \end{align*}

    The above inequality leads to $\hat R_1^t \le 1.5 R_1^t$, a contradiction.
\end{proof}

\cut{
\begin{figure*}[t!]
  \centering
  \subfigure[Number of users] {
    \includegraphics[width = 0.22\textwidth]{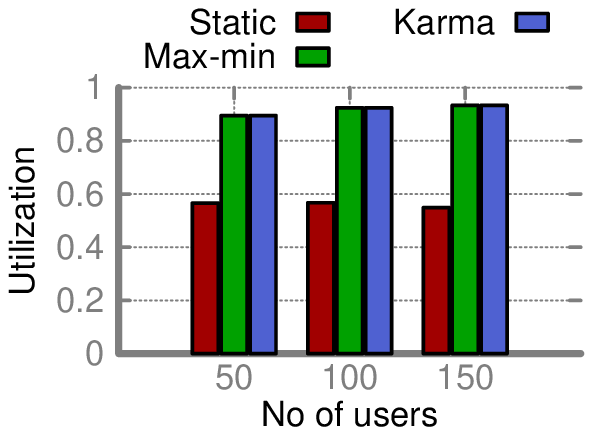}
    \label{fig:varyn-util}\hspace{2em}
    \includegraphics[width = 0.22\textwidth]{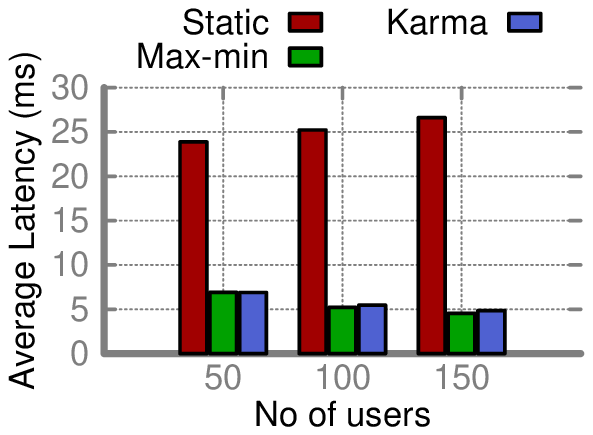}
    \label{fig:varyn-perf}\hspace{2em}
    \includegraphics[width = 0.22\textwidth]{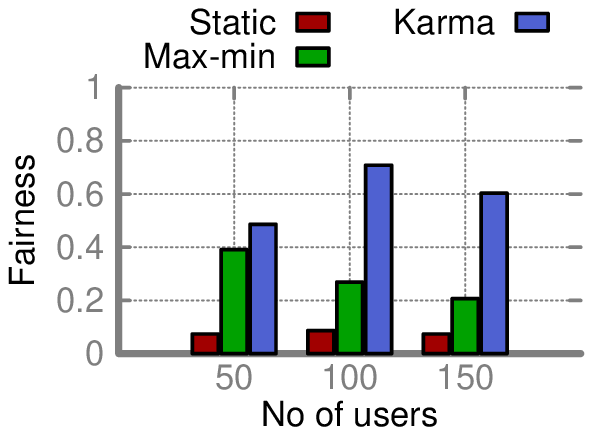}
    \label{fig:varyn-fairness}
  }
  \subfigure[Slice size] {
    \includegraphics[width = 0.22\textwidth]{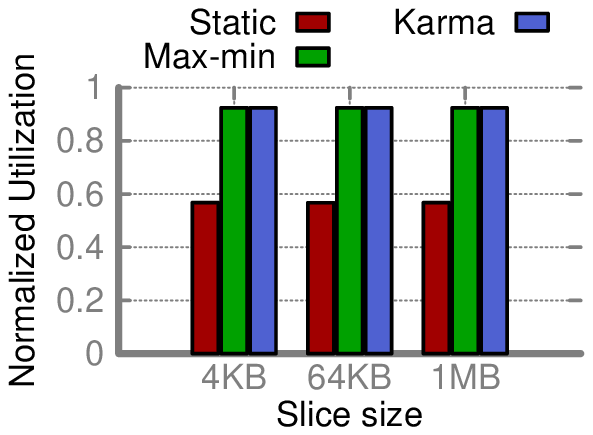}
    \label{fig:varyb-util}\hspace{2em}
    \includegraphics[width = 0.22\textwidth]{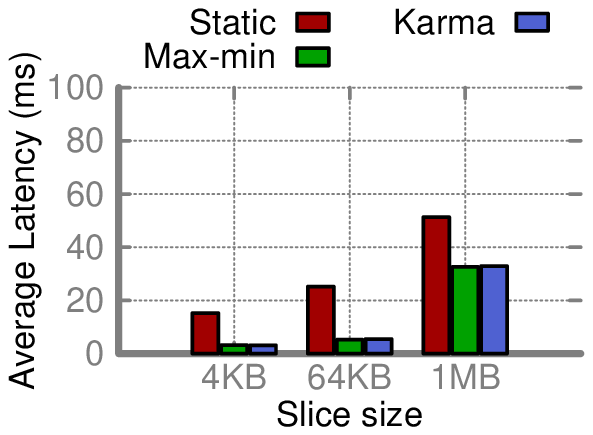}
    \label{fig:varyb-perf}\hspace{2em}
    \includegraphics[width = 0.22\textwidth]{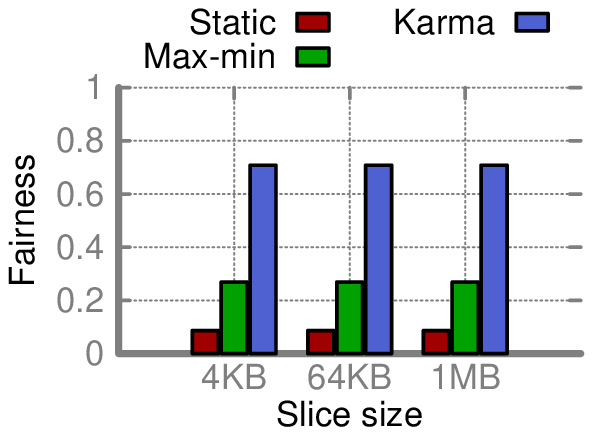}
    \label{fig:varyb-fairness}
  }%
  \vspace{-1em}
  \caption{\textbf{\name sensitivity analysis.} Variation of \name utilization and performance with (a) number of users and (b) slice size.} %
  \label{fig:sensitivity-pc}
  \label{fig:sensitivity-nb}
\end{figure*}
}

We now prove that if users in \name misreport their demand, they might lose a large amount of useful resources.

\begin{proof}[Proof of Lemma 2 -- decrease in useful allocation]
    Consider $n$ users and $f=1$. We will examine the first 3 quanta of two scenarios. In the first scenario, only users A, B, and C have non-zero demands, meaning we can ignore the other users. In Table \ref{tab:ex1} we see the demands of the users, as well as the resources they get allocated if they are truthful.

    \begin{table}[h]
    \centering
    \begin{tabular}{c|lll}
        t & 1                   & 2                    & 3 \\ \hline
        A & $n$ ($\frac{n}{2}$) & $n$ ($\frac{n}{4}$)  & $n$ ($\frac{3n}{8}$) \\
        B & $n$ ($\frac{n}{2}$) & $0$ ($0$)            & $n$ ($\frac{5n}{8}$) \\
        C & $0$ ($0$)           & $n$ ($\frac{3n}{4}$) & $0$ ($0$) \\
        others & $0$ ($0$)      & $0$ ($0$)            & $0$ ($0$) 
    \end{tabular}
    \vspace{1em}
    \caption{Example where it is in A's best interest to lie by misreporting a 0 demand on the first quantum. Each cell has the demand of the user and the parentheses contain the resources allocated if every user is truthful.}
    \label{tab:ex1}
    \end{table}

    It turns out that it is in A's best interest to lie on the first quantum and report a demand of $0$. This leads to A getting $\frac{5n}{4}$ resources, compared to $\frac{9n}{8}$, had A been truthful.

    Now we examine the second scenario in Table \ref{tab:ex2}, where every user other than A has the same demand.

    \begin{table}[h]
    \centering
    \begin{tabular}{c|lll}
        t      & 1         & 2                     & 3 \\ \hline
        A      & $n$ ($n$) & $n$ ($1$)             & $n$ ($1$) \\
        others & $0$ ($0$) & $1$ ($1$)             & $1$ ($1$)
    \end{tabular}
    \vspace{1em}
    \caption{Example where if A misreports its demand according to the demands of Table \ref{tab:ex1}, they lose resources. Each cell has the demand of the user and the parentheses contain the resources allocated if every user is truthful.}
    \label{tab:ex2}
    \end{table}

    Comparing Tables \ref{tab:ex1} and \ref{tab:ex2} we notice that A has the same demand for all quanta. This means that if A decides to misreport $0$ demand on the first quantum, in order to maximize its resources according to the first scenario, A would lose resources in the second one. More specifically, in the second scenario misreporting leads to A getting only $2$ resources, compared to $n+2$ resources for truthful reporting, i.e. misreporting is a factor of $\frac{n+2}{2}$ worst.
\end{proof}

\subsection{Generalized Algorithm}
\label{appendix:weighted-algorithm}

Algorithm~\ref{algo:karma-weighted-algorithm} is a generalized version of \name for users with different fair shares.

For the proof of Theorem~\ref{thm:collusion} and the proofs of the guarantees we prove for Algorithm~\ref{algo:karma-weighted-algorithm} we refer the reader to \cite{fikioris2021incentives}, where we conduct a deeper theoretical analysis of resource allocation under dynamic demands. More specifically, we prove that in Algorithm~\ref{algo:karma-weighted-algorithm}, colluding users cannot increase their collective useful allocation by over-reporting their demand. Additionally, we prove that no group of colluding users can increase their total useful allocation by a factor larger than $2$ by under-reporting demand.

\begin{algorithm}[!t]
  \small
  \caption{
      {\bf: \name with different fair shares.}
  }
  \label{algo:karma-weighted-algorithm}
  {\tt demand[u]}: demand of user \texttt{u} in the current quantum\\
  {\tt credits[u]}: credits of user \texttt{u}  in the current quantum\\
{\tt alloc[u]}: allocation of user \texttt{u}  in the current quantum\\
  {$f_u$}: fair share of user \texttt{u}\\
  {$w_u$}: normalized weight of user \texttt{u} $\left(\frac{f_u}{\sum_k f_k}\right)$\\
  {$\alpha$}: guaranteed fraction of fair share\\
  \\
  Every quantum do:
\begin{algorithmic}[1]
\State \texttt{shared\_slices} $\gets$ $(1 - \alpha) \cdot (\sum_k f_k)$
\State For each user {\tt u}, 
\State \hskip2em increment {\tt credits[u]} by $(1-\alpha)\cdot \frac{\sum_k f_k}{n}$
\State \hskip2em {\tt donated\_slices[u]} $=$ {\tt max} (0, $\alpha \cdot f_k -$ {\tt demand[u]})
\State \hskip2em {\tt alloc[u]} $=$ {\tt min} ({\tt demand[u]}, {$\alpha \cdot f_k$})
\State {\tt donors} $\gets$ all users {\tt u} with {\tt donated\_slices[u]} $> 0$
\State {\tt borrowers} $\gets$ all users {\tt u} with\\\hskip3em {\tt alloc[u]}$<${\tt demand[u]} \& {\tt credits[u]}>0
\vspace{0.1in}
\While{{\tt borrowers} $\neq\phi$ and \\\hskip3em ($\sum_{u}$ {\tt donated\_slices[u]} > 0 or \texttt{shared\_slices} > 0)}
  \State $b^\star$ $\gets$ borrower with maximum {\tt credit}s
  \If{{\tt donors} $\neq \phi$}
    \State $d^\star$ $\gets$ donor with minimum {\tt credit}s
    \State Increment {\tt credits}[$d^\star$] by $1$ 
  \State Decrement {\tt donated\_slices[u]} by $1$ 
  \State Update the set of {\tt donors} (line $6$)
  \Else
    \State Decrement {\tt shared\_slices} by $1$
  \EndIf
  \State Increment user $b^\star$ allocation by $1$
  \State Decrement {\tt credits}[$b^\star$] by $\frac{1}{n\cdot w_{b^\star}}$
  \State Update the set of {\tt borrowers} (line $7$)
 \EndWhile

\end{algorithmic}
\end{algorithm}

}

\end{document}